\UseRawInputEncoding
\documentclass[fleqn,usenatbib]{mnras}

\usepackage{graphicx}	
\usepackage{amsmath}	
\usepackage{amssymb}	
\usepackage{multicol}        
\usepackage{bm}		
\usepackage{pdflscape}	
\usepackage{hyperref}
\usepackage{url}

\newcommand{\angstrom}{\textup{\AA}}
\newcommand{\ms}{M_{\star}}

\usepackage[T1]{fontenc}
\usepackage{ae,aecompl}

\usepackage{newtxtext,newtxmath}

\title[A study of 14 structures behind Holm15A]{A spectroscopic study of 14 structures behind Holm15A: Detecting a galaxy group candidate at z=0.58.}
\author[H. Ibarra-Medel]{H. Ibarra-Medel$^{1}$\thanks{Contact e-mail: \href{mailto:hjibarram@gmail.com}{hjibarram@gmail.com}}
\\
$^{1}$Instituto de Astronom\'ia y Ciencias Planetarias, Universidad de Atacama, Copayapu 485, Copiap\'o, Chile}

\begin{document}
\label{firstpage}
\pagerange{\pageref{firstpage}--\pageref{lastpage}}
\maketitle

\begin{abstract}
Holm15A hosts one of the most massive back holes ever known. Hence, it is important to characterize any structure within its core to avoid any wrong association with its central black hole and, therefore, bias any future study. In this work, we present the first identification and characterization of 14 structures hidden behind the surface brightness of Holm15A. We model and subtract the spectral contribution of Holm15A to obtain the spectral information of these structures. We spectroscopically confirm that the 14 objects found are not associated with Holm15A. Ten objects have a well-defined galaxy spectrum from which we implement a fossil record analysis to reconstruct their past evolution. Nine objects are candidates members to be part of a compact galaxy group at redshift 0.5814. We find past mutual interaction among the group candidates that support the scenario of mutual crossings. Furthermore, the fossil reconstruction of the group candidates brings evidence that at least three different merger trees could assemble the galaxy group. We characterize the properties of the galaxy group from which we estimate a lower limit of the scale and mass of this group. We obtain a scale of $>$146$\pm$3 kpc with a dispersion velocity of 622$\pm$300 km/s. These estimations consider the lensing effects of the gravitational potential of Holm15A. The other five objects were studied individually. We use public archive data of integral field spectroscopic observations from the Multi-Unit Spectroscopic Explorer instrument.
\end{abstract}


\section{Introduction}

Holm15A can be considered one of the most interesting objects in the sky, since its well resolved core structure permits the detailed study of the nature of its supermassive black hole (SMBH). Holm15A is the brightest cluster galaxy (BCG) of Abell 85 \citep{Abell+1989} with a redshift of z=0.055 \citep{Lopez-Cruz+2014}. Its surface brightness profile has one of the largest cores ever observed that supports the idea that it contains one of the largest SMBH\footnote{\citet{Mehrgan+2019} estimate an SMBH mass of $4\times 10^{10}M_{\odot}$.} ever found \citep{Lopez-Cruz+2014,Mehrgan+2019}. However, the nature of the SMBH of Holm15A is not completely understood. It is still an open question the existence of a single, a dual or more complex system of SMBHs. Because of this, it is vital to characterize all the observed structures within Holm15A to disentangle the scenario that better describes the nuclear region, particularly, the number of SMBH. In this study, we explore, classify, dissect and analyze a set of 14 observed astronomical objects that were hidden within the surface brightness of Holm 15A. 

For this aim, we use integral field spectroscopy (IFS) observations with the Multi-Unit Spectroscopic Explorer (MUSE) instrument. MUSE is located at the Very Large Telescope (VLT) of the European Southern Observatory \citep[ESO,][]{Bacon+2017}. The IFS observational techniques spatially resolves the spectral information of an observed target through the use of integral field units (IFU) devices \citep[see the review of ][]{Sanchez+2020}. With the spatially resolved spectral information, the IFS technique has improved our ability to study the extragalactic physics and spatially dissect the galaxy properties within a global and local view \citep{Ibarra-Medel+16}. 

On the other hand, we also use the stellar population synthesis to study the IFS data. The stellar populations synthesis (or archaeological reconstruction) are a strong tool to understanding the galaxy evolution \citep[e.g,][]{Tinsley+1980,Buzzoni+1989,Bruzual+2003,Kauffmann+2003May,Kauffmann+2003ab,Cid-Fernandes+2005,Gallazzi+2005,Tojeiro+2007,Walcher+2011,Goddard+2017}. These techniques, also known as the fossil record, model the observed  galaxy spectra energy distribution (SED) as the assembly of multiple single stellar populations (SSP) with different ages and metallicities \citep{Conroy+2013}. The fossil record method resolves the inverse problem to find the flux contribution of each SSP to the observed spectrum. With this information, it is possible to reconstruct the past evolution of the star formation and enrichment of galaxies throughout the cosmic times \citep{Perez+2013,Artemi+2022,Zhou+2022}. However, the fossil record has the limitation to only access the histories at the sky positions at the observed time \citep{Ibarra-Medel:2019aa}. 

With the advent of large IFU surveys and state of the art IFU observations \citep[e.g.,][]{Bacon+2001,Cappellari+2011,Croom+2012,Sanchez+2012,Cid-Fernandes+2013,Bundy+2015}, now it is possible to apply the fossil record method to the resolved spectral information of the galaxies and dissect the galaxy formation through time and space \citep[e.g.,][]{Ibarra-Medel+16,Gonzalez-Delgado+2016,Gonzalez-Delgado+2017,Garcia-Benito+2017,Ibarra-Medel:2019aa,Peterken+2020,Peterken+2021,Ibarra-Medel+2022}. In this point, \citet{Lopez-Cruz+2019} applied the stellar population synthesis to disentangle the star formation histories of four galaxy members of the Seyfert’s Sextet (HCG 79). In that study, they show that it is possible to apply the fossil record to trace the past interactions of galaxy systems by looking into common star formation episodes \citep{Hickson+1992,Plauchu-Frayn+2012,Moura+2020}. Hence, the stellar population synthesis is a powerful tool to understand the nature of the galactic systems, uncover their past evolution and demonstrate the past inter-linkage among members of more complex gravitational systems like galaxy groups.

The outline of the paper is arranged as follows. In Section \ref{methods}, we describe the observational data that where used for this study, and describe the tools used to perform the stellar population synthesis. In Section \ref{analisis}, we explain our spectroscopic analysis, describing the methods and steps to fit the surface brightness of Holm15A. We describe how the spectra of the 14 detected objects where extracted and describe all the steps for the fossil reconstruction. In Section \ref{sp_results}, we describe the spectroscopic results of all objects, from which we conclude the existence of a possible galaxy group behind Holm15A. In Section \ref{group_ana}, we present a complete characterization of the galaxy group candidate. In Section \ref{discuss}, we give a discussion of the results, and present our conclusions. We have adopted a cosmology with $\Omega_m=0.27$, $\Omega_{\lambda}=0.72$ and $H_0=71\ kms^{-1}Mpc^{−1}$.

\section{Data and Methods}\label{methods}

\begin{figure*}
	\includegraphics[width=0.45\textwidth]{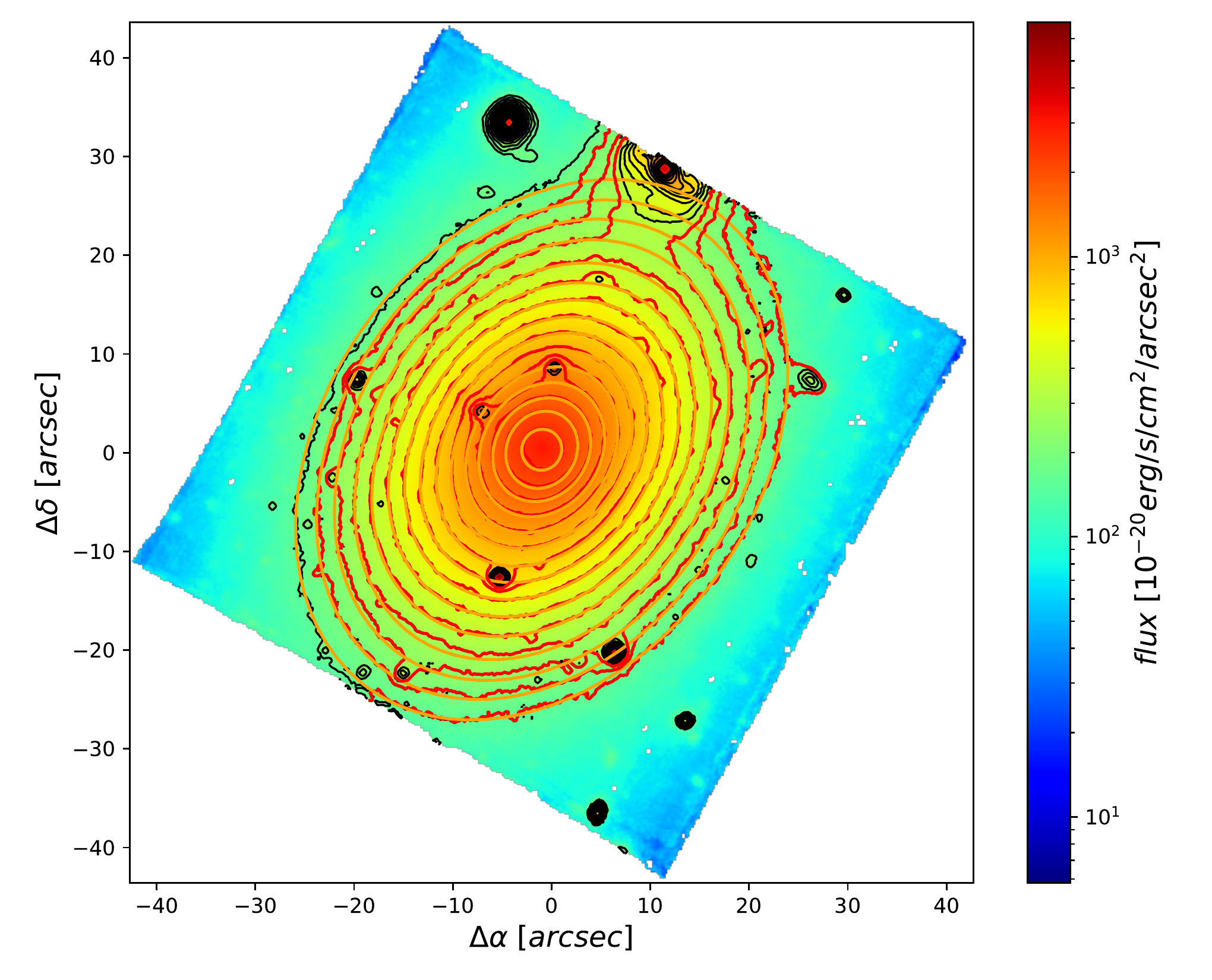}
	\includegraphics[width=0.45\textwidth]{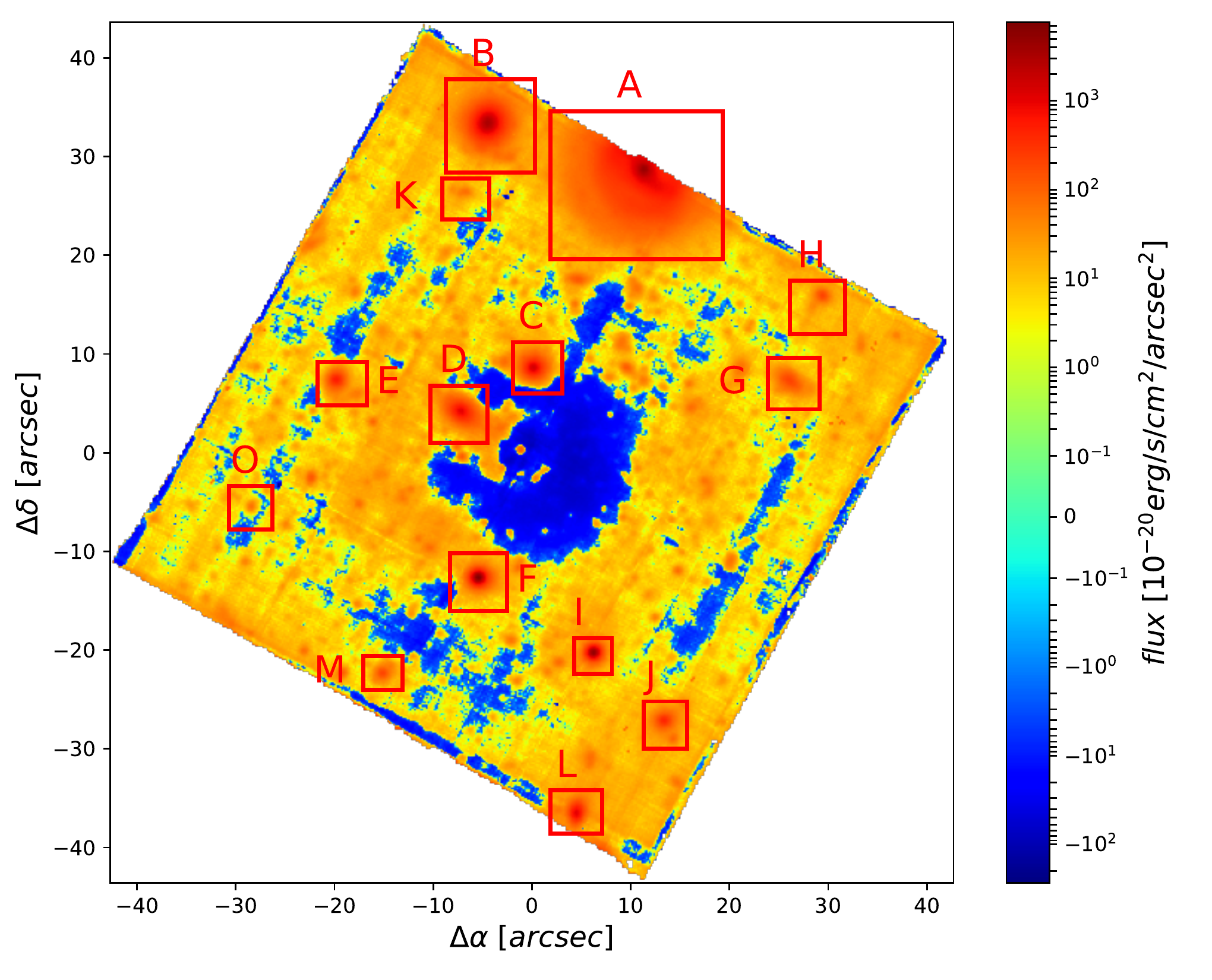}
    \caption{Left panel: Collapsed SB map of Holm15A from $4750-9300\angstrom$. The black lines represent the full set of SB isophotes within the MUSE FoV. The red line represent the selected isophotes to fit the ellipse model. The orange lines represent the fitted ellipse model. The color bar represents the observed SB flux. Right panel: The residual map of the subtracted 2D SB model of Holm15A. The red thick boxes represent the extracted subsets for the 14 detected objects. The color bar represents the observed SB flux.}
    \label{fig:Muse_map}
\end{figure*}

\subsection{MUSE Data}

The MUSE instrument \citep{Bacon+2017} in its Wide Field Mode (WFM) provides a field of view of $59".9\times60"$ with a spatial sampling of $0".2$ per pixel. The average spatial resolution of the MUSE IFS has a full-width at half maximum (FWHM) between $0".3$ to  $0".5$. However, in practice, the spatial resolution is limited by seeing. The MUSE spectral range covers the optical region from $4750\angstrom$ (blue) to $9300\angstrom$ (red), with spectral resolving power of 1770 (blue) to 3590 (red). For this study, we use the archive data taken with the MUSE WFM of Holm15A with the program ID 099.B-0193\footnote{\url{https://archive.eso.org/wdb/wdb/eso/eso_archive_main/query?prog_id=099.B-0193}} \citep{Mehrgan+2019}. The observation was taken on the 16th of November 2017 with the data set ID MUSE.2017-11-16T03:50:01.938. While \citet{Mehrgan+2019} measured an FWHM spatial resolution of $0".72$ on the red part of the spectrum, we measured an FHWM of $0".63$ at $7000\angstrom$ and $0".58$ at $9000\angstrom$ using the two point sources on the field\footnote{The reported seeing on the weather logs at Paranal during the observation night was $0".47$, see \url{https://www.eso.org/asm/ui/ambient-server?site=paranal&mjd=58073.159744650000000&exptime=1260.000&mark=MUSE.2017-11-16T03:50:01.938&color=orange}}. We use a Moffat profile \citep{Trujillo+2001} to model the FWHM of the observed seeing within the MUSE datacube. A detailed description of the seeing modeling is given on Appendix\ref{psf_ana}.

\subsection{Spectral Fitting}

For this work, we use the {\sc pyPipe3D}\footnote{\url{https://gitlab.com/pipe3d/pyPipe3D/}} code that uses the {\sc pyFIT3D} pipeline \citep[][]{Lacerda+2022}, which it is based on the previous Pipe3D/FIT3D code \citep[][]{Sanchez16a,Sanchez16b}. The workflow of {\sc pyPipe3D} can be summarized as follows: 
\begin{itemize}
\item[a)]  It extracts an integrated central spectrum to estimate the initial value of the redshift and kinematics. 
\item[b)]  It performs a spatial binning across the FoV of the input datacube to target a signal-to-noise ratio (SNR) of 50 per bin. The SNR is defined within a spectral window that the user can modify \citep{Lacerda+2022}. 
\item[c)] {\sc pyFIT3D} performs a full stellar population synthesis (SPS) on the coadded spectra within each spatial bin. In this paper, {\sc pyFIT3D} fits the stellar continuum by using the updated version of the single stellar population (SSP) models of {\sc Bruzual \& Charlot 03} \citep[Bruzual et al. 2022 in prep, Sanchez et al. 2022 in prep, ][]{Mejia2022} that uses the \verb|MaSTAR| stellar library \citep{MaSTARS}. This library uses a \citet{Salpeter+55} initial mass function (IMF) and it is composed by a grid of $N_A$ ages $\times$ 7 metallicities\footnote{The SSP libraries are available at \url{http://ifs.astroscu.unam.mx/pyPipe3D/templates/}}. In this study, we use three SSP grids, with $N_A=$ 39, 36 and 34 ages and have a maximum age of 13.5, 8.5 and 6.25 Gyr due to the observed redshift of our objects (see Section \ref{sp_results}). The SSP ages are approximately linearly spaced below $0.02$ Gyr and then it is logarithmically spaced (Sanchez et al. 2022 in prep). The SSP metallicities cover the values: $Z_\star$=0.0001, 0.0005, 0.0020, 0.0080, 0.0170, 0.0300, 0.0400, with $Z_{\odot}=0.019$. For the stellar synthesis, {\sc pyFIT3d} considers a \citet{Cardelli+1989} dust attenuation law.
\item[d)]  {\sc pyPipe3D} saves the results of the spectral inversion done by {\sc pyFIT3D} and reconstructs a set of 2D maps that contains the stellar population properties. During this process, {\sc pyPipe3D} undone the initial segmentation binning and creates a set of 2D maps that contains the same number of spaxels of the original datacube. For the case of the SPS decomposition, {\sc pyPipe3D} generates a map of the weights for each SSP per each age and metallicity grid. With these weights, we can reconstruct the star formation histories and the accumulative stellar mass growth across the time \citep{Ibarra-Medel+16,Ibarra-Medel+2022}. 
\item[e)]  Finally, {\sc pyPipe3D} performs the nebular emission analysis, providing fluxes and kinematic maps of the emission lines (Sanchez 2022 et al. in prep). 
\end{itemize}

In \citet{Ibarra-Medel:2019aa}, the {\sc Pipe3D} code was fully tested using hydro-dynamical cosmological simulations and MaNGA \citep{Bundy+2015} type IFU mock observations. In that work, it was concluded that the code effectively reconstructs the qualitative evolution of the galaxies with the fossil record method. At younger ages (from the observed time), the code quantitatively recovers the evolution, but at older ages ($\gtrsim 6 Gyr$), the reconstruction is valid in a qualitative way.

\begin{figure}
	\includegraphics[width=\columnwidth]{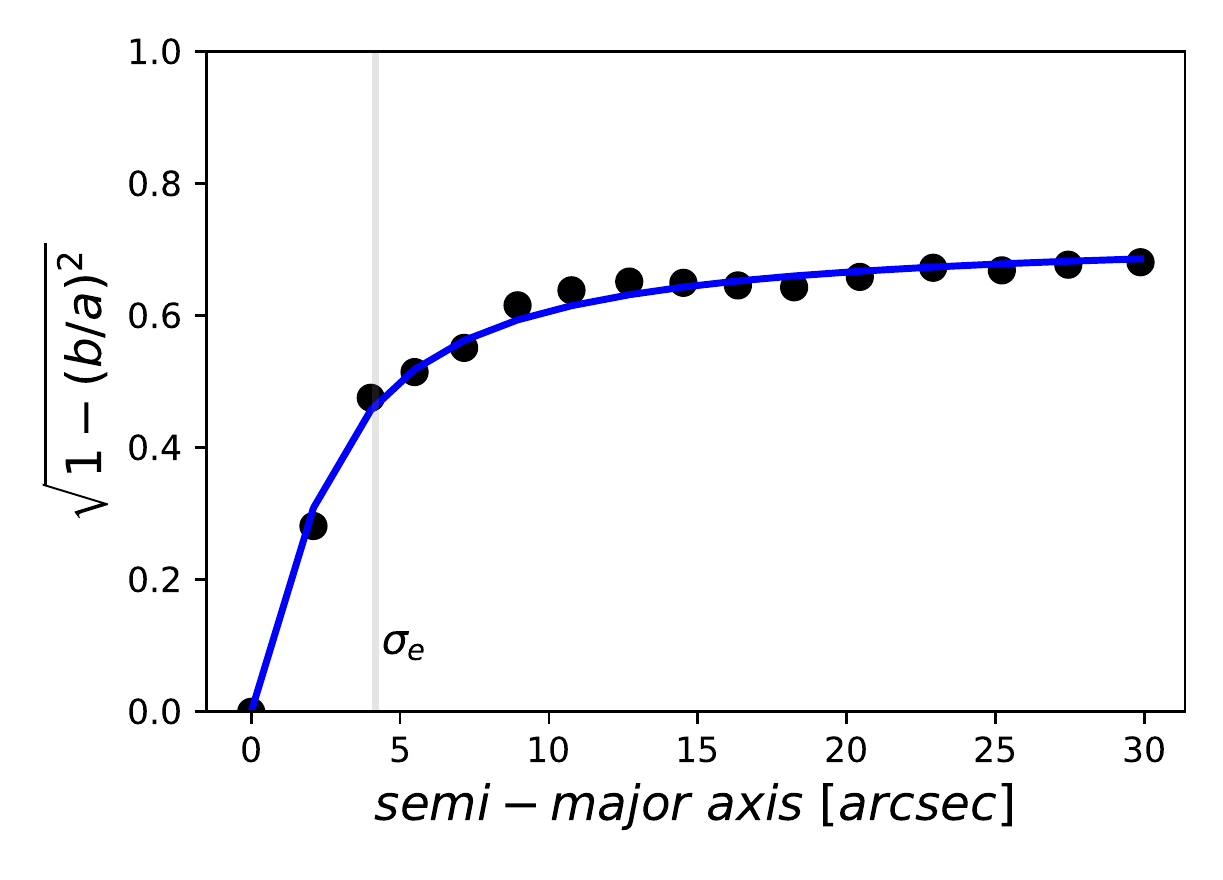}
	\includegraphics[width=\columnwidth]{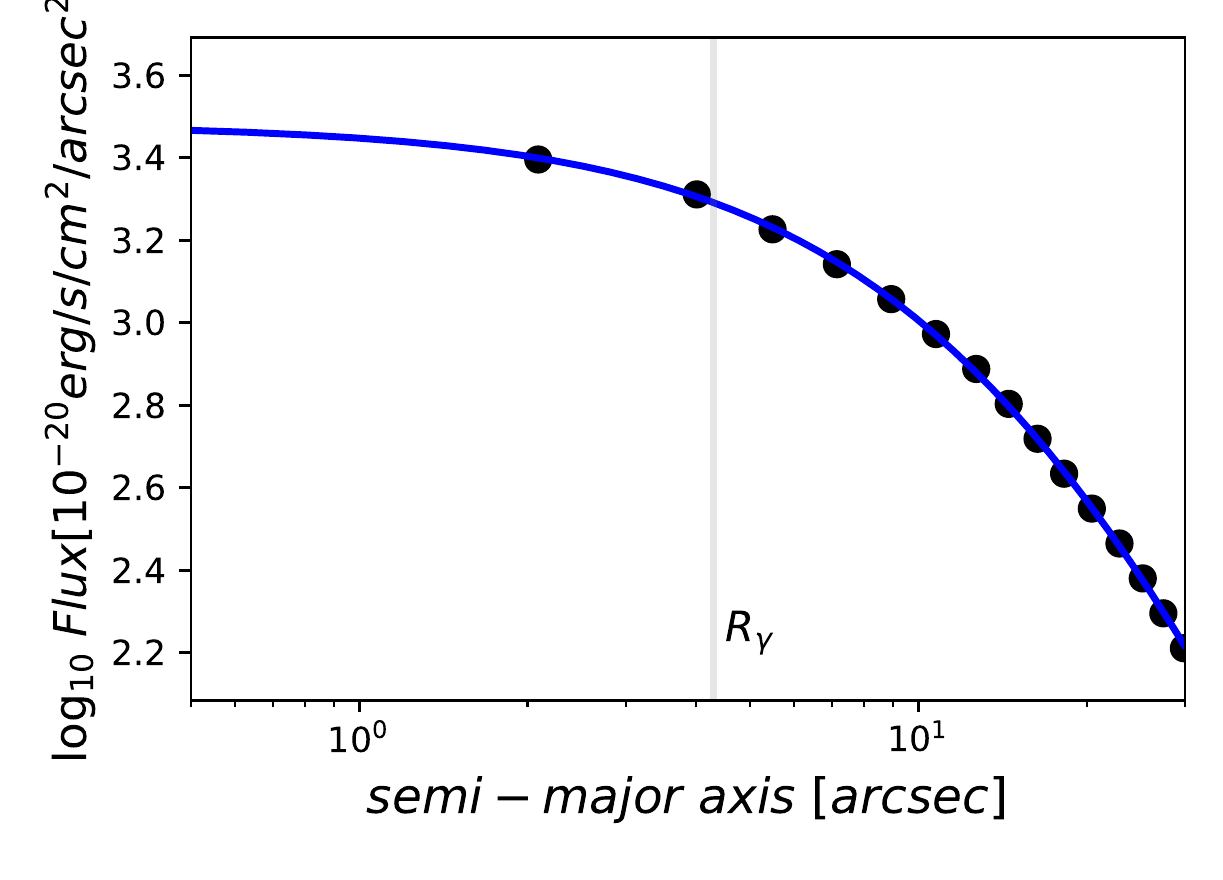}
    \caption{Upper panel: The measured eccentricity profile of Holm15A in terms of the semi-major axis. The blue line represents the best fitted model. The gray vertical line represents the value of $\sigma_e$, that is the braking point of the observed circularization of the issophotes. Lower panel: The measured SB profile of Holm15A in terms of the semi-major axis. The blue line represents the best fitted model using a Nuker profile. The gray vertical line represents the value of $R_{\gamma}$, that is the cusp radius.}
    \label{fig:fit_prof}
\end{figure}

\section{Analysis}\label{analisis}

\subsection{Holm15A Surface Brightness Modeling}\label{sb_model}

Our principal objective is to analyze, model, and characterize the astronomical objects hidden by the brightness of Holm15A. To perform this task, we first correct the MUSE datacube by the galactic extinction with the use of the \citet{Schlafly+2011} galactic dust extinction maps and with the \citet{Fitzpatrick+1999} extinction law. After this correction, we model the surface brightness (SB) distribution of Holm15A across the FoV of the MUSE WFM. For this purpose, we first run {\sc pyPipe3} over the full datacube to obtain the cube of the stellar model of Holm15A. Then, we integrate the stellar model cube through their full spectral axis ($4750-9300\angstrom$) to obtain a high SNR collapsed image. We fit the SB isophotes using ellipses with the zero-point, semi-mayor axis ($a$), semi-minor axis ($b$), and rotation angle ($\Phi$) as free parameters. We use 15 flux logarithmic spaced isophotes covering all the FoV along the major axis of Holm15A. We show these isophotes in the left panel of Figure~\ref{fig:Muse_map} with their corresponding fitted ellipses. 
Then, to generate a first model of the SB profile of Holm15A, we parameterize the eccentricity (defined as $e=\sqrt{1-(b/a)^2}$) profile in terms of the semi-mayor axis $a$. The parameterization can be well represented by:

\begin{equation}
    e(a)=\frac{2A_e}{\pi}\arctan\left(\frac{a\pi}{2\sigma_e}\right).
\end{equation}

Figure \ref{fig:fit_prof} shows the best fit of the eccentricity profile with $A_e=0.73$ and $\sigma_e=4".45$ ($4.77$ kpc at $z=0.056$). The value of $\sigma_e$ marks the brake point where the circularization of the isophotes ends, and a constant eccentricity value dominates the eccentricity profile. For the rotation angle $\Phi$, we define a constant $\Phi$ profile to model the SB distribution. For that objective, we obtain the median value of $\Phi$ of all the fitted ellipses with a major-axis value larger than $2\sigma_e$. The final value of $\Phi$ is $-34^{\circ}.3\pm0^{\circ}.9$.

We then use the isophotes levels to fit the SB profile with a Nuker profile \citep{Lauer+1995} defined as:

\begin{equation}
I(r)=2^{(\beta-\gamma)/\alpha}I_0\left(R_b/r\right)^{\gamma}\left[1+(r/R_b)^{\alpha}\right]^{(\gamma-\beta)/\alpha}. \label{nuker}
\end{equation} Defining the cusp radius as $R_{\gamma}=R_b\left(\frac{1/2-\gamma}{\beta-1/2}\right)^{1/\alpha}$ \citep{Lopez-Cruz+2014}, with $R_b$ as the brake radius. The cusp radius defines the scale where $dlog I/dlog r =-1/2$ and indicates the region where the SB profile becomes shallow \citep{Mehrgan+2019}. In addition, \citet{Mehrgan+2019} found that Holm15A has a no defined brake radius. Therefore, we will use the cusp radius $R_{\gamma}$ as the core scale of Holm15A. In a forthcoming paper (Lopez-Cruz et al., in prep), we will explore in detail the profile of Holm15A.

To minimize the fitting space parameter of the SB fit, we use the fitted exponential values of the Nuker profile reported by \citet{Lopez-Cruz+2014}, leaving $I_0$ and $R_b$ as a free parameters. The values reported by \citet{Lopez-Cruz+2014} are $\alpha=1.24$, $\beta=3.33$ and  $\gamma=0.0$. We find that the best fit values to model the SB profile of Holm15A with a Nuker profile are $I_0=4.69$ $10^{-18}erg/s/cm^2/arcsec^2$ with a $R_{\gamma}$ value of $4".31$ ($4.63$ kpc). For comparison, \citet{Lopez-Cruz+2014} found a $R_{\gamma}=4".26$ while \citet{Mehrgan+2019} found a $R_{\gamma}=4."1$. Hence, our measurement of $R_{\gamma}$ is consistent with both estimations. Figure~\ref{fig:fit_prof} presents the best fit for the SB profile.

Therefore, we can reconstruct a two-dimension SB model with the fitted values of $A_e$, $\sigma_e$, $I_o$ and $\Phi$. After this step, we obtain a high-SNR collapsed flux map (integrating over the spectral axis) from the original MUSE datacube. By subtracting the SB model map from the flux collapsed map, we obtain a high-SNR two-dimensional residual map (see the right panel of Figure \ref{fig:Muse_map}). 

Then, we generate a three-dimensional SB cube model by fitting the SB distribution through all the spectral sampling of the original MUSE cube. To fit the SB models, we fix the value of $\Phi$ and fit the eccentricity and SB profiles (as described above) for each of the 3,682 spectral values of the MUSE spectral axis. Hence, we obtain an SB model cube with the exact spectral sampling of the original MUSE observation. Finally, we obtain a residual cube by subtracting the SB model cube from the original flux cube. The final residual cube contains all the spectral information of the astronomical objects without the stellar spectra of Holm15A.

\subsubsection{Substructure detection}

The high-SNR 2D residual map of Holm15A reveals the existence of multiple objects behind Holm15A. 
We select 14 structures listed in Table~\ref{tab1} and Figure~\ref{fig:Muse_map}. The selection criteria require that the structures have a spectrum with an SNR larger than two or at least three prominent nebular emission lines to identify its redshift value. Once we select the structures, we define a set of extraction areas to obtain subsets of the residual spectral datacube as individual datacubes. In addition, from the residual datacube, we select four regions without any visible structure to obtain an average sky spectrum. We use this sky spectrum to correct the extracted datacubes by any sky emission still present in the MUSE datacube. We show all the extraction regions in Figure~\ref{fig:Muse_map}, each region was defined to contain all the resolved spectral information of each object. Then, we perform a simple SB fit for each object to obtain their principal structural parameters, such as the effective radii ($R_e$), position angle $\Phi$, and central coordinates. For this fit, we follow the same procedure described in Section \ref{sb_model} by using a \citet{Sersic1963} profile. We list these measurements in Table~\ref{tab1}.

\begin{table}
\begin{center}
\caption{List of the 14 objects found behind the SB profile of Holm15A.}\label{tab1}
\resizebox{0.5\textwidth}{!}{
\begin{tabular*}{0.6\textwidth}{@{\extracolsep{\fill}} c c c c c c c c c c }
\hline\hline
ID & R.A & Dec. & $z$ & $R_e$ & $R_e$ & $\Phi$ & Type \\
   & [FK5]    & [FK5]    &     & [arcsec] & [kpc] & [deg] &  \\
\hline 
A & $0:41:49.70$ & $-9:17:42.57$ & 0.0498 & $3".09$ & 2.96 & $-28.8$ & Gal \\
B & $0:41:50.76$ & $-9:17:37.80$ & 0.0563 & $0".93$ & 1.00 & $-25.1$ & Gal \\
C & $0:41:50.45$ & $-9:18:02.57$ & 0.5857 & $0".62$ & 4.09 & $ -9.3$ & Gal \\
D & $0:41:50.95$ & $-9:18:06.95$ & 0.5832 & $1".37$ & 9.03 & $-33.0$ & Gal \\
E & $0:41:51.80$ & $-9:18:03.79$ & 0.5766 & $0".59$ & 3.85 & $  7.9$ & Gal \\
F & $0:41:50.83$ & $-9:18:23.97$ & 1.5637 &         &      &         & Qso \\
G & $0:41:48.70$ & $-9:18:03.89$ & 0.5834 & $1".06$ & 6.97 & $-36.8$ & Gal \\
H & $0:41:48.48$ & $-9:17:55.30$ & 0.5827 & $0".87$ & 5.72 & $-32.6$ & Gal \\
I & $0:41:50.04$ & $-9:18:31.47$ & 0.000  &         &      &         & Str \\
J & $0:41:49.55$ & $-9:18:38.29$ & 0.5769 & $0".79$ & 5.17 & $ 23.1$ & Gal \\
K & $0:41:50.93$ & $-9:17:44.85$ & 0.5832 &         &      &         & Gal \\
L & $0:41:50.16$ & $-9:18:47.69$ & 0.5757 & $0".94$ & 6.14 & $-13.1$ & Gal \\
M & $0:41:51.48$ & $-9:18:33.49$ & 1.0126 & $0".97$ & 7.86 & $ 39.9$ & Gal \\
0 & $0:41:52.38$ & $-9:18:16.78$ & 0.5831 &         &      &         & Gal \\
\hline
\end{tabular*}}
\end{center}
\end{table}

\begin{figure*}
	\includegraphics[width=\textwidth]{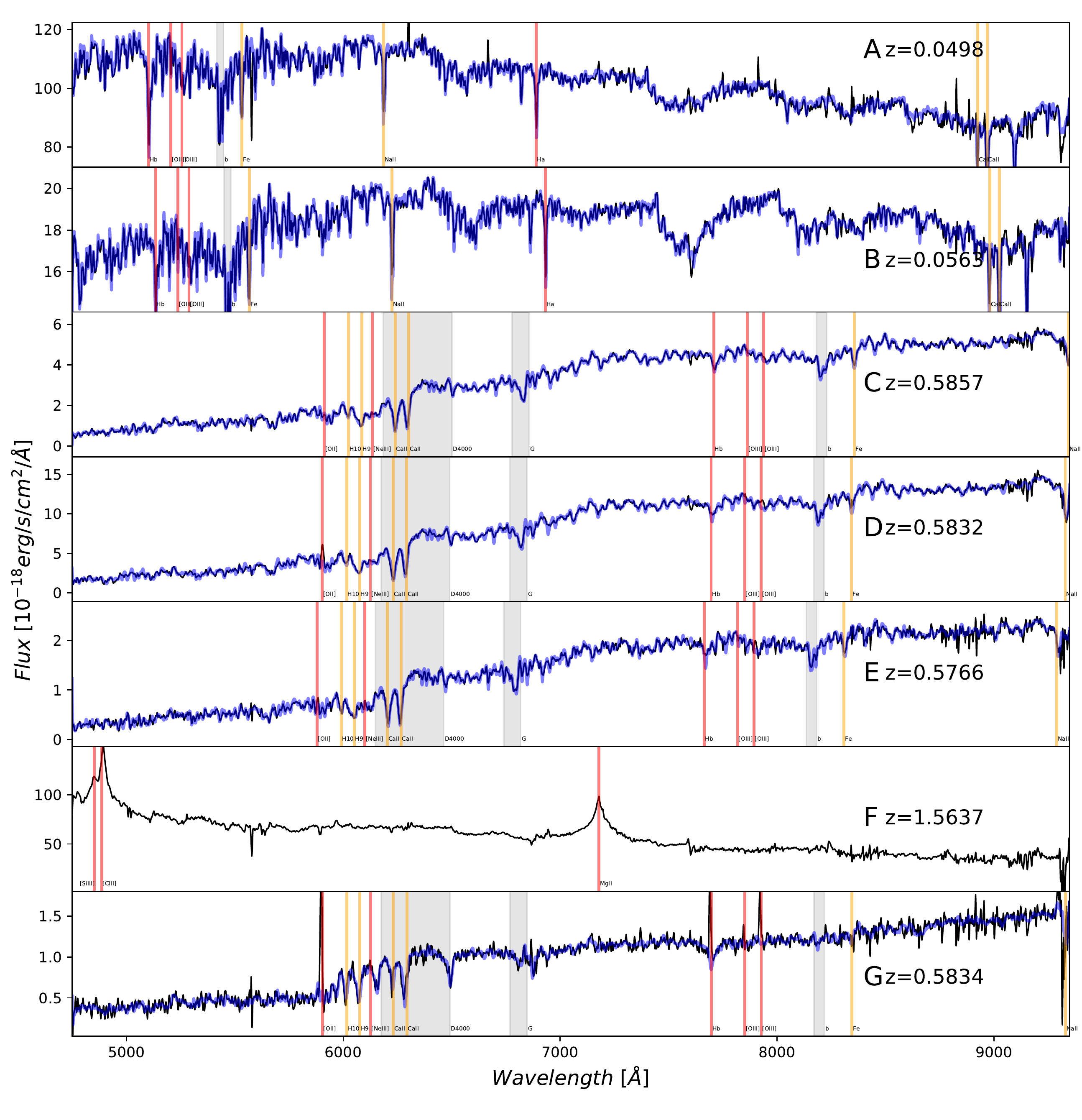}
	\caption{Central spectra at the observed wavelength within 0.5$R_e$ for objects \textbf{A}, \textbf{B}, \textbf{C}, \textbf{D}, \textbf{E} and \textbf{G}. For object \textbf{F} we plot the total spectrum within their region. The blue line in each spectrum represent the fitted {\sc PyFIT3D} stellar model. The vertical red lines represent the most prominent emission  nebular lines, the vertical yellow lines represent the most prominent stellar absorption lines. The gray bands represent the {\sc D4000} index region, and the Fraunhofer G and b bands.}
	\label{plot_spectra2}
\end{figure*}

\begin{figure*}
	\includegraphics[width=\textwidth]{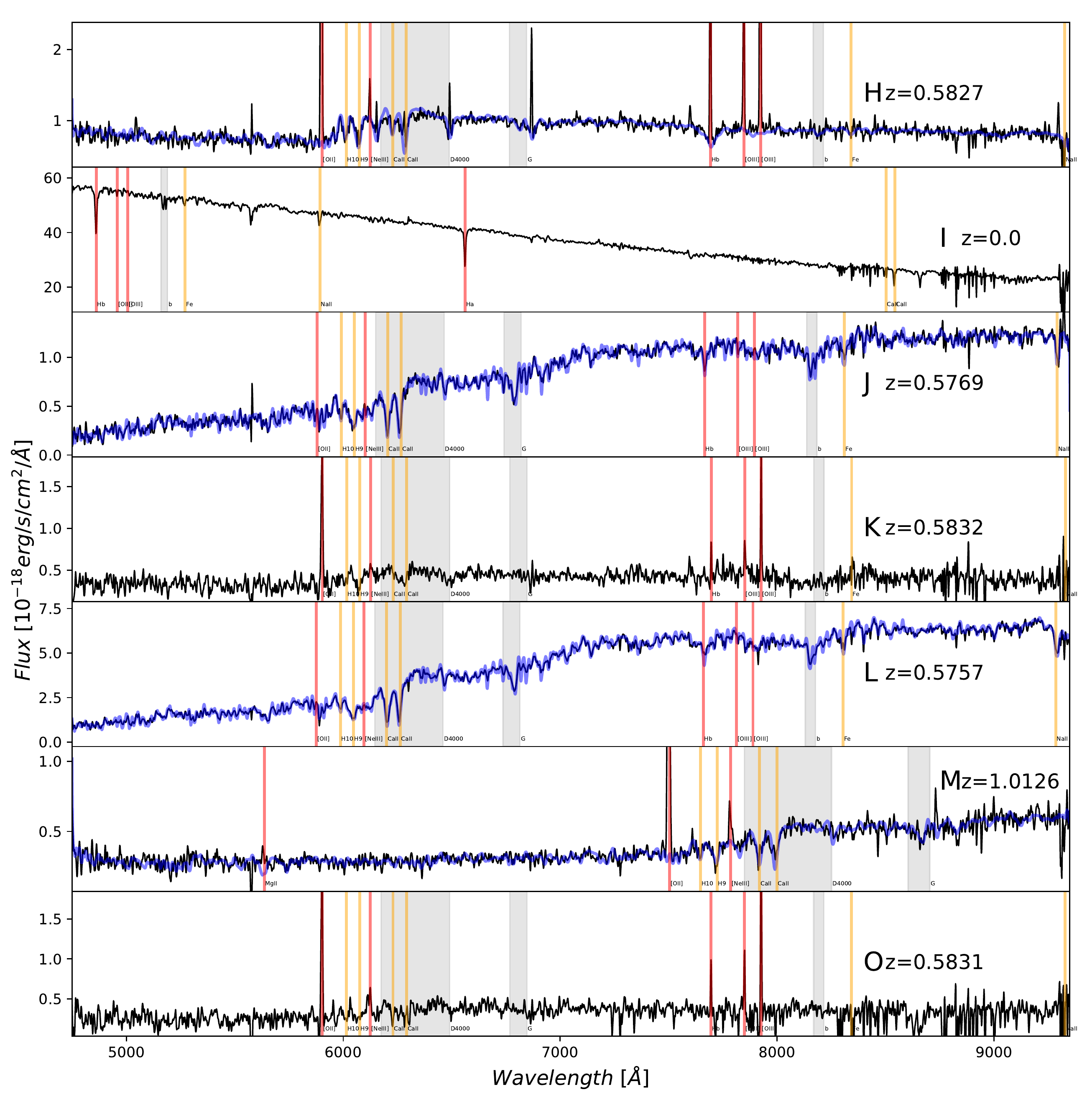}
	\caption{Central spectra at the observed wavelength within 0.5$R_e$ for objects \textbf{H}, \textbf{J}, \textbf{K}, \textbf{L}, \textbf{M} and \textbf{O}. For object \textbf{I} we plot the total spectrum within their region. The blue line in each spectrum represent the fitted {\sc PyFIT3D} stellar model. The vertical red lines represent the most prominent emission  nebular lines, the vertical yellow lines represent the most prominent stellar absorption lines. The gray bands represent the {\sc D4000} index region, and the Fraunhofer G and b bands.}
	\label{plot_spectra3}
\end{figure*}

\subsection{Spectral Analysis.}

By visually inspecting the spectra of each object, we are able to classify them into three types: galaxies, quasars, and stars (see Table~\ref{tab1}). We determine that objects $\mathbf{F}$ and $\mathbf{I}$ are a quasar and a field star respectively. We measure the redshift of the quasar in object $\mathbf{F}$ by identifying the {Mg\sc II}$\lambda2800$, {C\sc III]}$\lambda1909$ and {Si\sc III]}$\lambda1892$ emission lines. Those lines are typically strong emission lines in quasars \citep{Vanden+2001}. Due to the nature of the objects, we discard the star spectrum of the object $\mathbf{I}$, and we separate analysis the quasar spectrum of the object $\mathbf{F}$ in Appendix~\ref{qsr_t}. We cannot detect a stellar spectrum for objects $\mathbf{K}$ and $\mathbf{O}$ but we can identify an intense emission for the nebular lines of {[O\sc II]}$\lambda3728$, $H_{\beta}$,  and {[O\sc III]}$\lambda\lambda 4959,5007$.  Due to the absence of a well defined stellar spectrum, we cannot apply the full spectral analysis (see below), and therefore we can only use the nebular emission lines to measure their redshifts.

For the remaining objects, we proceed to fully analyze their spectra by using the {\sc pyPIPE3D} code. For illustrative purposes, we plotted all the central spectra for our objects in Figure~\ref{plot_spectra2} and ~\ref{plot_spectra3}. We extract the central spectra within $0.5R_e$, and over-plotted the fitted stellar spectra model from {\sc pyPIPE3D}. With the SPS of {\sc pyFIT3D}, we can apply a fossil record reconstruction to study the galaxy evolution across the cosmic time. We follow the methodologies described in \citet{Ibarra-Medel+16,Lopez-Cruz+2019,Ibarra-Medel+2022} and in \citet{Lacerna+20} to reconstruct their specific star formation histories (sSFH). We briefly describe these methodologies below:

With the SSP decomposition obtained by {\sc pyPIPE3D}, we use the SSP weights ($f_{ssp}$) to estimate the instantaneous star formation rate (SFR) at a given SSP age. The history of the instantaneous SFR at a given age is the star formation history (SFH), and it is defined as follows:

\begin{equation}
\begin{split}
    SFH(t_i)=\sum_j f_{ssp}(Z_j,t_i)\times L_n\times10^{0.4A_V}\\ \times \frac{\Upsilon_{\star}(Z_j,t_i)}{m_{loss}(Z_j,t_i)\Delta_{ti}},
\end{split}
\end{equation} where $f_{ssp}(Z_j,t_i)$ is the SSP weight for the metallicity $Z_j$ and age $t_i$. The value of $L_n$ represents the normalized flux used during the SSP analysis. In this case, we define the normalized flux as the integrated flux within $4500-5500\angstrom$ in the rest-frame of each galaxy \citep[see][]{Lacerda+2022}. For the object $\mathbf{M}$, we require to redefine this window as $3850-4250\angstrom$ due to its high redshift. $A_v$ is the extinction value obtained by the SSP fit, $\Upsilon_{\star}(Z_j,t_i)$ is the mass-to-ligt ratio of the SSP with metallicity $Z_j$ and age $t_i$. $m_{loss}(Z_j,t_i)$ is the mass-loss factor of the SSP with age $t_i$ and metallicity $Z_j$. Therefore, the instantaneous mass growth that happens at age $t_i$ is $f_{ssp}(Z_j,t_i)\times L_n\times10^{0.4A_V}\times \Upsilon_{\star}(Z_j,t_i)/m_{loss}(Z_j,t_i)$. $\Delta_{ti}$ is the characteristic time gap between the SSP age $t_{i-1}$ and $t_{i+1}$. The ratio of these two quantities defines the SFR at age $t_i$. Finally, we can define the observed time SFR rate derived from the stellar population synthesis ($SFR_{ssp}$) as the average value of $SFH(t)$ within the last $20$ $Myr$.

Similarly, we are able to reconstruct the accumulative stellar mass (ASM) distribution across the time as:

\begin{equation}
\begin{split}
    \ms(>t)=\sum_i^{t\le t_{i}}\sum_j f_{ssp}(Z_j,t_i)\times L_n\times10^{0.4A_V}\\ \times \Upsilon_{\star}(Z_j,t_i).
\end{split}
\end{equation}

The main difference between the ASM and the accumulative SFH(t) is the correction of the mass-loss factor of each SSP. The ASM did not correct the mass-loss and always had a monotonic and positive increment. In addition, we calculate the formation ages at which the ASM reaches its $90\%$ ($T_{90}$) and $50\%$ ($T_{90}$) of the observed time total stellar mass. Finally, we define the specif SFH (sSFH) as:
\begin{equation}
    sSFH(t)=\frac{SFH(t)}{\ms(>t)}
\end{equation}

\subsubsection{Single Stellar Population Properties}

We also obtain the average age and metallicity values for the stellar populations weighted by light ($lw$) and mass ($mw$). We use the definitions of $lw$ and $mw$ used in \citet{Sanchez16b,Ibarra-Medel:2019aa,Ibarra-Medel+2022,Lacerna+20}, and in Cano-Diaz et. al 2022 (in prep). These mass/light weighted quantities are defined as follows:
\begin{equation}
\log(X_{mw})=\frac{\sum_i^{t\le t_{i}}\sum_j\log(X_{j,i})f_{ssp}(Z_j,t_i)\times \Upsilon_{\star}(Z_j,t_i)}{\sum_i^{t\le t_{i}}\sum_j f_{ssp}(Z_j,t_i)\times \Upsilon_{\star}(Z_j,t_i)},
\label{eq_ageMW}
\end{equation}
\begin{equation}
\log(X_{lw})=\frac{\sum_i^{t\le t_{i}}\sum_j\log(X_{j,i})f_{ssp}(Z_j,t_i)}{\sum_i^{t\le t_{i}}\sum_j f_{ssp}(Z_j,t_i)},
\label{eq_ageLW}
\end{equation} where $X_{j,i}$ is the quantity at the grid with age $t_i$ and metallicity $Z_j$. 

Finally, to have a model independent estimation of the age of the stellar populations, we use the {\sc pyPipe3D} integrated {\sc D4000} index. The {\sc D4000} index is defined as the ratio between the continuum flux within the $4050-4250\angstrom$ and $3750-3950\angstrom$ spectral windows \citep{Bruzual83,Sanchez16b}. The above is a sanity check: the {\sc D4000} index is an index that only depends on the spectra and it increases with older stellar population \citep[e.g.,][]{Poggianti+97,Kauffmann+2003May}. 

\begin{figure}
    \includegraphics[width=\columnwidth]{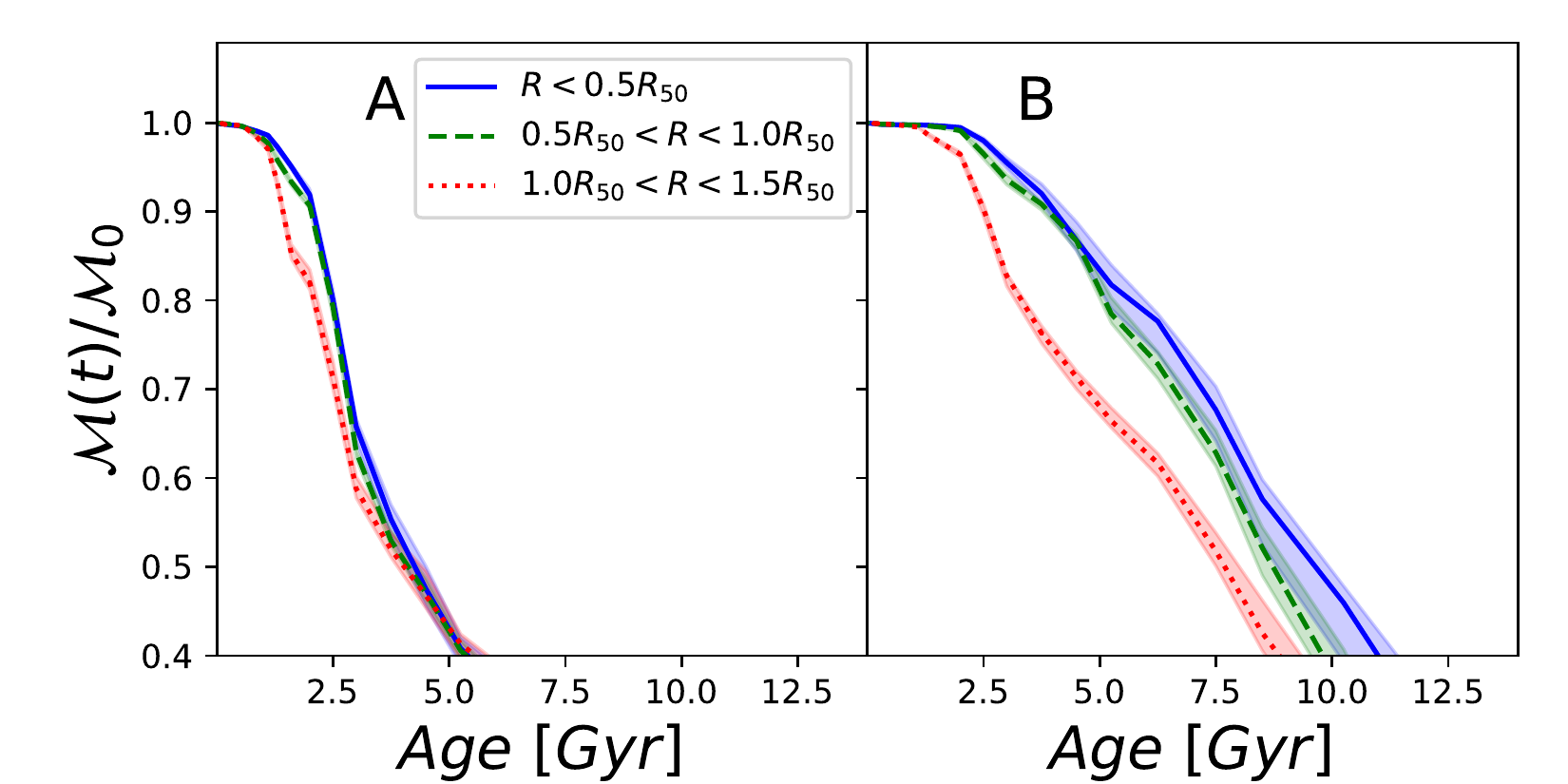} \includegraphics[width=\columnwidth]{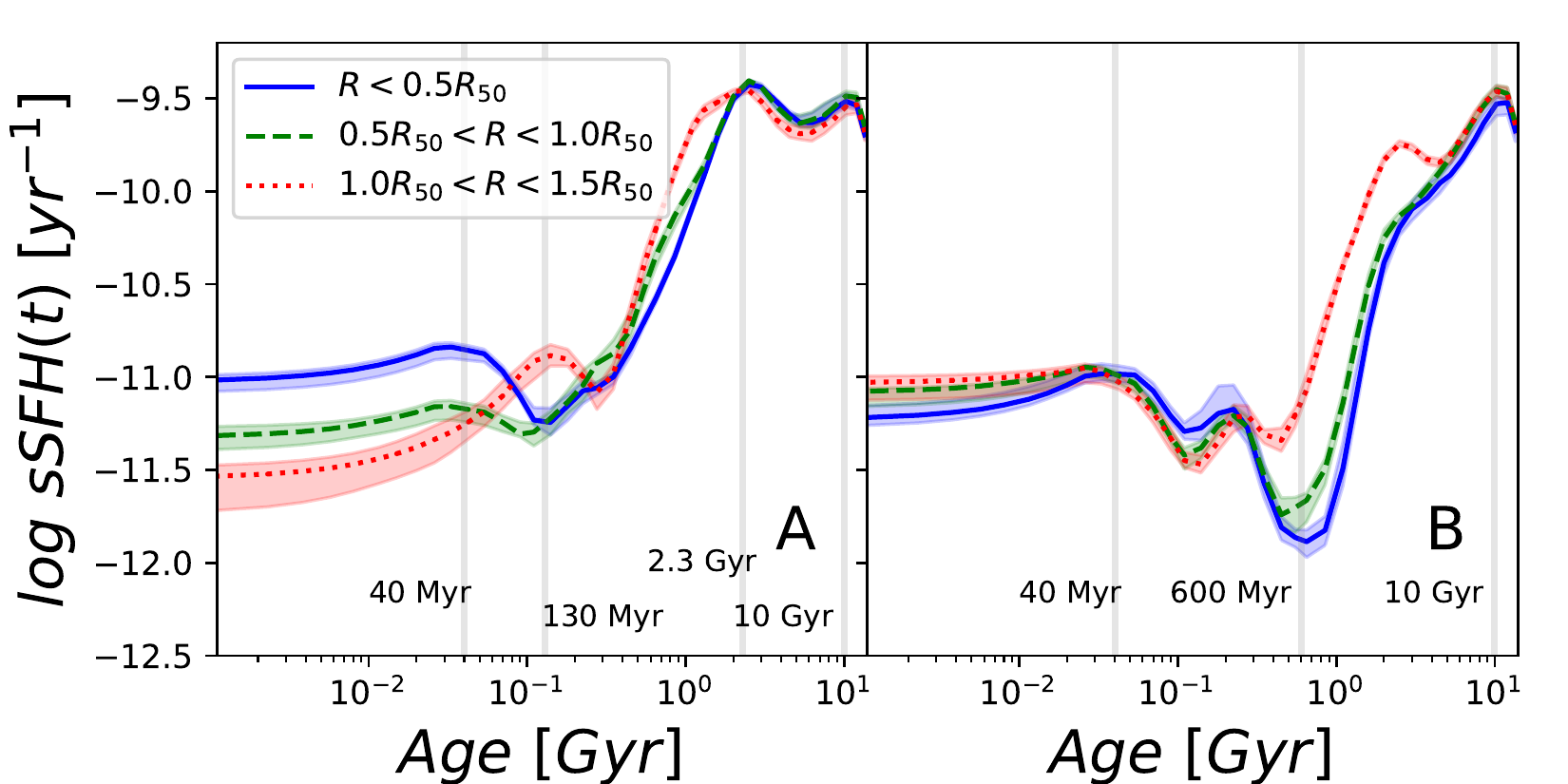} \caption{Upper panels: The ASM distributions for objects $\mathbf{A}$ and  $\mathbf{B}$, at three radial regions: $R < 0.5R_{50}$ (blue), $0.5R_{50} < R < R_{50}$ (green), and $R_{50} < R < 1.5R_{50}$ (red). Lower panels: The sSFH for object \textbf{A} and \textbf{B} at the same radial bins and the same color code of the upper panel. The shaded color areas are the $1\sigma$ variance for 100 iterations. The vertical gray lines represent the SF burst and quenching episodes described in the text.}
    \label{A_B_SFH}
\end{figure}

\begin{figure*}
	\begin{tabular}{cc}
	\includegraphics[width=0.47\textwidth]{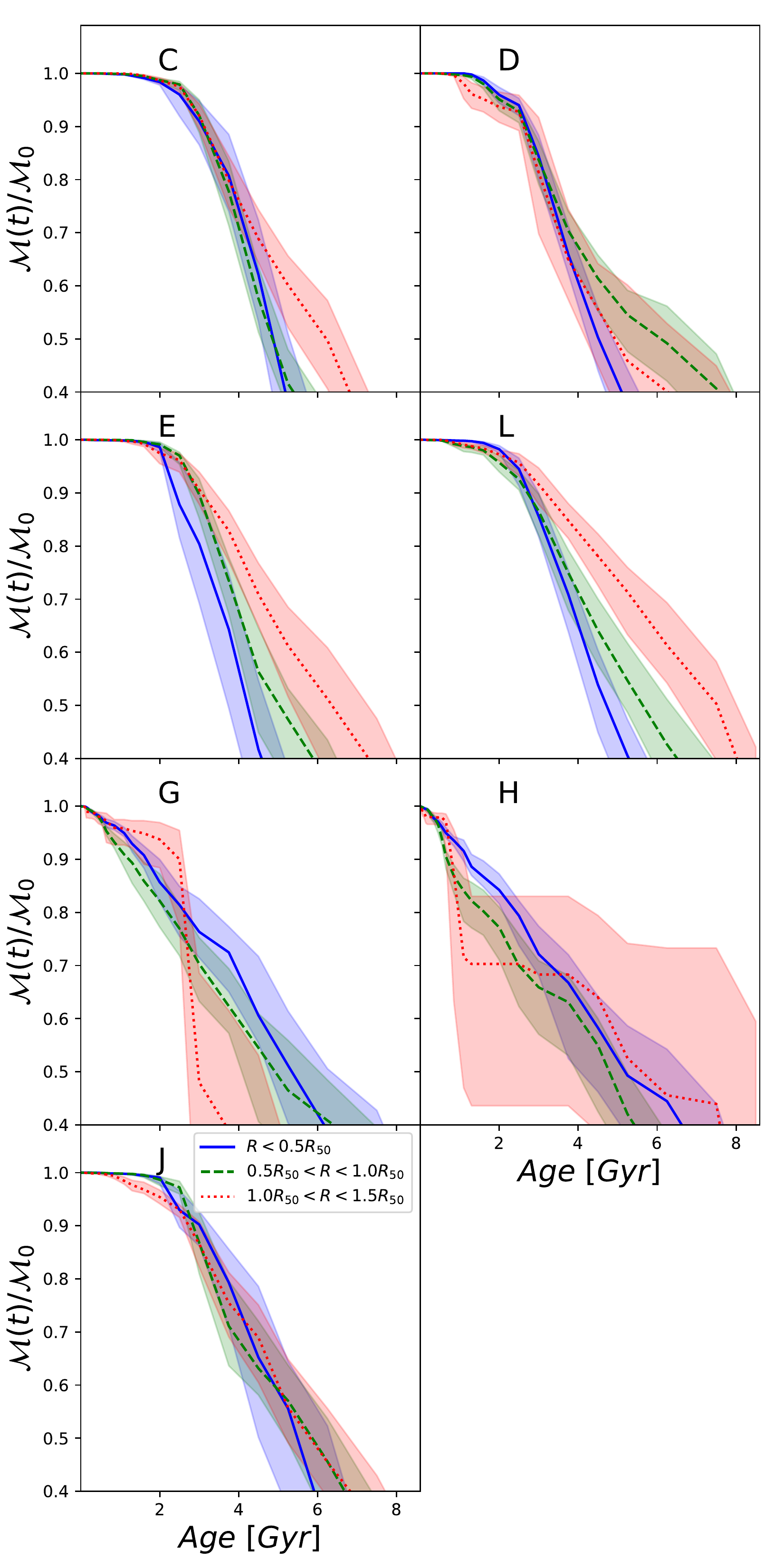} &
	\includegraphics[width=0.47\textwidth]{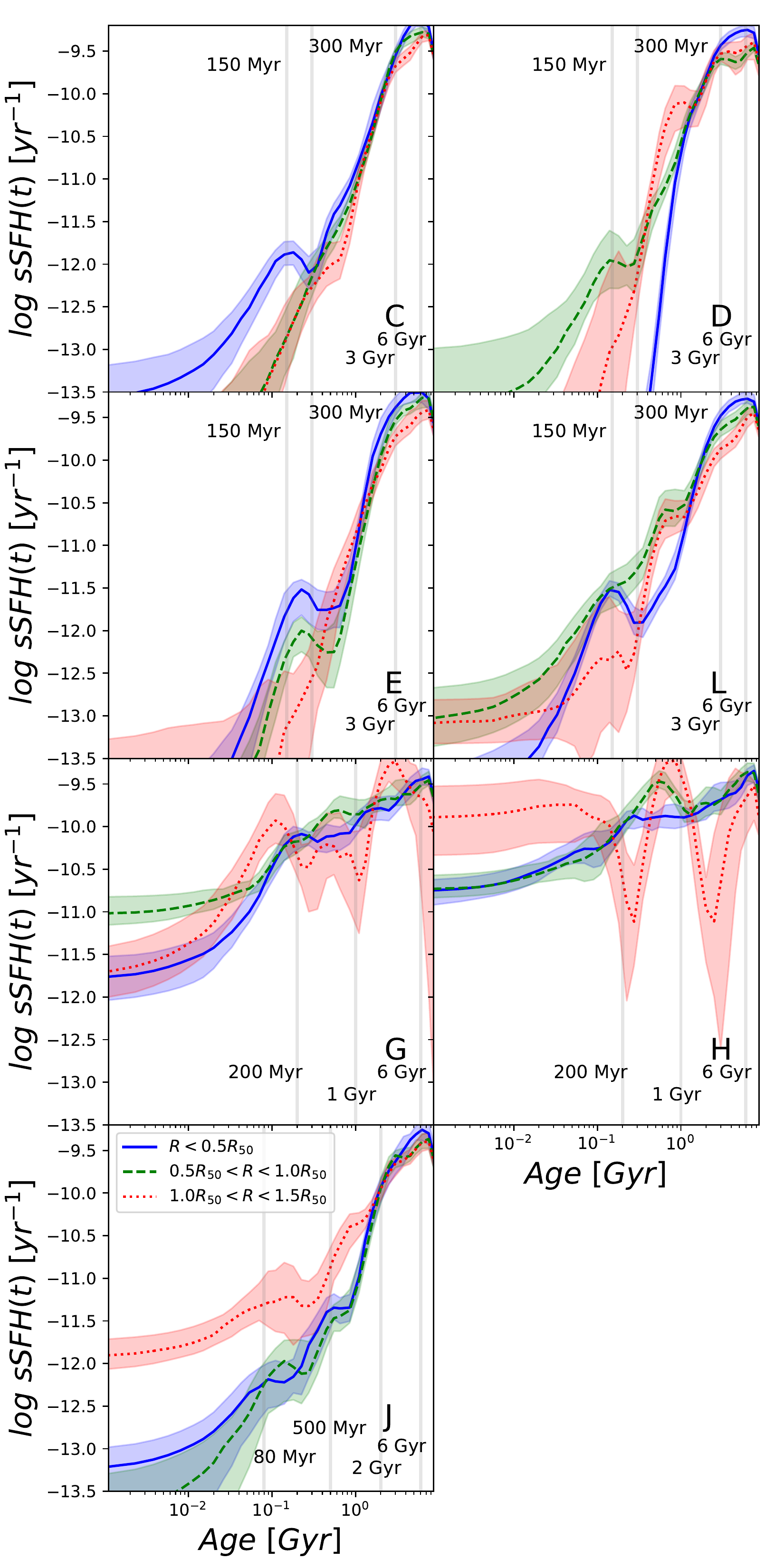} \\
	\end{tabular}
    \caption{Left panels: The ASM distributions for objects \textbf{C}, \textbf{D}, \textbf{E}, \textbf{G}, \textbf{H}, \textbf{J} and \textbf{L} at three radial regions: $R < 0.5R_{50}$ (blue), $0.5R_{50} < R < R_{50}$ (green), and $R_{50} < R < 1.5R_{50}$ (red). Right panels: The sSFH for the same objects at the same radial bins and the same color code of the left panels. The shaded color areas are the $1\sigma$ variance for 100 iterations. The vertical gray lines represent the SF burst and quenching episodes described in the text. Objects \textbf{C}, \textbf{D}, \textbf{E} and \textbf{L} are classify as interacted objects, \textbf{G} and \textbf{H} are SF objects and \textbf{J} is classify as intermediate object.}
    \label{C_L_SFH}
\end{figure*}

\begin{table*}
\begin{center}
\caption{The SSP properties from the SPA and spectral analysis within one effective radii.}\label{tab2}
\resizebox{\textwidth}{!}{
\begin{tabular*}{0.95\textwidth}{@{\extracolsep{\fill}} c c c c c c c c c c c }
\hline\hline
ID & $\ms$ & {\sc D4000}$^a$ & $Age_{lw}$ & $Age_{mw}$ &  $\log Z_{lw}^c$ & $\log Z_{mw}^c$ & $\log sSFR_{ssp}^b$ & $T_{50}$ & $T_{90}$ \\
  & $[M_{\odot}]$ &  & $[Gyr]$ & $[Gyr]$ & $[Z/Z_{\odot}]$ & $[Z/Z_{\odot}]$ &  & $[Gyr]$ & $[Gyr]$ \\
\hline 
A & $5.8\pm0.1\times10^{9}$  &                & $1.5\pm 0.1$ &  $4.1\pm 0.1$ & $-0.06\pm 0.01$ & $-0.01\pm 0.01$ & $-11.1\pm 0.1$ & $4.1\pm0.1$ & $2.1\pm 0.1$ \\
B & $3.7\pm0.1\times10^{9}$  &                & $2.9\pm 0.1$ &  $7.4\pm 0.2$ & $-0.24\pm 0.01$ & $-0.16\pm 0.01$ & $-11.1\pm 0.1$ & $8.7\pm0.3$ & $4.0\pm 0.1$ \\
C & $1.3\pm0.1\times10^{11}$ & $1.82\pm 0.01$ & $3.7\pm 0.1$ &  $4.5\pm 0.1$ & $+0.15\pm 0.02$ & $+0.17\pm 0.02$ & $-13.8\pm 0.2$ & $4.8\pm0.3$ & $3.1\pm 0.1$ \\
D & $4.3\pm0.7\times10^{11}$ & $1.89\pm 0.02$ & $3.5\pm 0.1$ &  $4.6\pm 0.2$ & $+0.07\pm 0.01$ & $+0.06\pm 0.02$ & $-13.7\pm 0.6$ & $5.1\pm0.5$ & $2.7\pm 0.1$ \\
E & $1.1\pm0.1\times10^{11}$ & $1.92\pm 0.02$ & $3.5\pm 0.2$ &  $4.4\pm 0.2$ & $+0.11\pm 0.03$ & $+0.07\pm 0.05$ &                & $5.0\pm0.2$ & $2.9\pm 0.2$ \\
G & $3.9\pm0.4\times10^{10}$ & $1.46\pm 0.02$ & $1.1\pm 0.1$ &  $3.9\pm 0.4$ & $-0.52\pm 0.07$ & $-0.32\pm 0.06$ & $-11.2\pm 0.2$ & $5.2\pm0.6$ & $1.5\pm 0.1$ \\
H & $1.1\pm0.3\times10^{10}$ & $1.30\pm 0.02$ & $0.6\pm 0.1$ &  $3.3\pm 0.3$ & $-0.59\pm 0.09$ & $-0.19\pm 0.07$ & $-10.7\pm 0.2$ & $4.8\pm0.5$ & $0.9\pm 0.2$ \\
J & $5.0\pm0.3\times10^{10}$ & $1.82\pm 0.02$ & $3.6\pm 0.2$ &  $4.6\pm 0.3$ & $-0.02\pm 0.03$ & $-0.01\pm 0.04$ & $-13.5\pm 0.3$ & $5.3\pm0.6$ & $2.8\pm 0.2$ \\
L & $2.6\pm0.2\times10^{11}$ & $1.85\pm 0.02$ & $3.4\pm 0.1$ &  $4.6\pm 0.2$ & $-0.00\pm 0.03$ & $+0.04\pm 0.05$ & $-13.1\pm 0.4$ & $5.3\pm0.3$ & $2.7\pm 0.1$ \\
M & $9.0\pm1.5\times10^{10}$ & $1.30\pm 0.02$ & $0.5\pm 0.1$ &  $4.0\pm 0.2$ & $-0.62\pm 0.08$ & $-0.23\pm 0.08$ & $-10.8\pm 0.1$ & $5.2\pm0.3$ & $2.4\pm 0.2$ \\
\hline
\end{tabular*}}
\end{center}
\raggedright
{ $^a$ The spectral coverage for objects $\mathbf{A}$ and $\mathbf{B}$ didn't reach the Balmer jump.}\\
{$^b$ Object $\mathbf{E}$, didn't returns a present day $sSFR_{ssp}$ value.}\\
{$^c$ $Z_{\odot}=0.019$.}\\
\end{table*}

\section{Spectral Results}\label{sp_results}
We separate all our objects into three groups due to their redshifts: The first group is formed by objects \textbf{A} and \textbf{B}, and their redshifts (z=0.0498,0.0563) identify them as a possible galaxy cluster members of Abell 85. At $z\approx$0.053, the age of the Universe is 12.9 Gyr. Hence, we use the SSP grid with 39 ages, with a maximum SSP age of 13.5 Gyr. The second group are objects \textbf{C}, \textbf{D}, \textbf{E}, \textbf{G}, \textbf{H}, \textbf{J}, \textbf{K}, \textbf{L} and \textbf{O}. Those objects have redshift values around 0.58, and are candidates to be a galaxy group behind Abell 85. At z=0.58, the Universe's age is 8.1 Gyr; therefore, we use the SSP grid with 36 ages with a maximum SSP age of 8.5 Gyr. The third group contains only the object \textbf{M} with a z=1.0126. The age of the Universe at z=1.0126 is 5.9 Gyr. Hence, we use the SSP grid with 34 ages with a maximum SSP age of 6.25 Gyr. At the redshift of z=1.0126, the Balmer jump is still visible within the spectral range of the MUSE instrument, and therefore it is possible to apply the fossil record method to the observed spectrum of object \textbf{M}. We listed in Table~\ref{tab2} all the final SSP properties for these objects.

\subsection{Object A and B}
The SB of object \textbf{A} shows evidence of a bar structure oriented along its mayor-axis. The spectral analysis shows that it is blue-shifted by 1,666 km/s from Holm15A. Taking into account that the reported velocity dispersion of Abell 85 is $\approx$ 750 km/s \citep{Bravo-Alfaro+2009,Lopez-Cruz+2014}, the line-of-sight relative velocity of object \textbf{A} indicates that it could be kinematically decoupled from Abell 85. The relative velocity with the central galaxy of Abell 85 is larger than two times the escape velocity of the cluster \citep{Serra+2011}. 

Figure~\ref{A_B_SFH} (left panel) shows the SPS reconstruction of the ASM and the sSFH of object \textbf{A} at three radial bins: $R<0.5R_e$, $0.5R_e<R<R_e$ and $R_e<R<1.5R_e$. The sSFH shows evidence of a constant star formation rate at all radii with two dominant bursts, one at the age of $\approx 10$ Gyr and a second at the age of $\approx 2.3$ Gyr. After that, the galaxy suffers a rapid star formation (SF) shutdown that begins at the central part of the galaxy. In the central region, another SF episode begins at $\approx 130$ Myr. This SF episode reactivates the SFR at the central region and has a peak ($\log sSFR\approx-11.7$ dex) at $\approx 40$ Myr. After this peak, the SF slowly decreases at its present value ($\log sSFR=-11.1$ dex). The ASM shows that the second burst at $\approx 2.3\ Gyr$ contributes to the assembly of at least half of the total mass of object $\mathbf{A}$. The assembly mode is almost uniform at all radii until the age of $\approx 2.5$ Gyr, where the inside-out assembly mode dominates. The $T_{50}$ age of object \textbf{A} is 4.1 Gyr, with a $T_{90}$ of 2.1 Gyr.

Object $\mathbf{B}$ did not show any evidence of a bar structure like the object $\mathbf{A}$. Its redshift shows that object $\mathbf{B}$ has a relative velocity with Holme15A of 280 km/s, implying that it is kinematically linked to the Abell 85 system. The fossil record reconstruction shows a similar but not equal sSFH as object $\mathbf{A}$. For object $\mathbf{B}$, we observe a first SF burst at late epochs ($\approx 10\ Gyr$) from which its sSFR declines, and it reaches a minimum at the age of $\approx 600\ Myr$. After this minimum, object $\mathbf{B}$ shows a new star forming burst that reaches its peak at $\approx 40\ Myr$ (see right panels of Figure~\ref{A_B_SFH}). The ASM shows a constant rate on the mass assembly until 2 Gyr, and it is dominated by an inside-out assembly mode \citep{Ibarra-Medel+16}. It has a $T_{50}$ of 8.7 Gyr and a $T_{90}$ of 4.0 Gyr.

Both objects have light-weighted ages older than 1.5 Gyr, with light-weighted metallicities below the solar value (see cols 4,5,6 and 7 of Table\ref{tab2}). The difference between the mass and light-weighted ages are 0.43 and 0.4 dex for objects $\mathbf{A}$ and $\mathbf{B}$. The difference between the mass and light-weighted metallicities are 0.06 and 0.07 dex each. The difference between the light and mass-weighted values traces the differences among the old and young stellar populations \citep{Panter+2008,Plauchu-Frayn+2012,Lacerna+20}. In this case, the differences among the light and mass metallicities for both objects give evidence of a possible rejuvenation due to the in-fall of pristine gas \citep{Maiolino+2019,Artemi+2022}. The in-fall of this gas could dilute the gas metallicity and feeds the last starburst observed on the sSFHs of objects $\mathbf{A}$ and $\mathbf{B}$.

\subsection{Objects C, D, E, G, H, J, K, L and O}\label{lens_group}

These objects are candidate members of a galaxy group with an average redshift of $0.5814\pm0.001$. We hereafter call this group as J004150-091812, and we present a complete analysis for the group in Section~\ref{group_ana}. In Figure~\ref{C_L_SFH}, we present the ASM and sSFH for each object of this group. From the sSFH, we can classify our objects into interacted objects, star-forming objects, and intermediate objects. 

\emph{Interacted objects}: This group consists of objects \textbf{C}, \textbf{D}, \textbf{E} and \textbf{L}. The SPS analysis of these objects return a set of sSFH's with evidence of past interactions. They have common SF burst at similar epochs among each other. In this aspect, \citet{Lopez-Cruz+2019} present evidence of a sheared star formation history among members of the Seyfert's Sextet galaxy group. The common SF bursts trace the epochs when the galaxies were crossing \citep{Hickson+1992,Plauchu-Frayn+2012}, and therefore, interacted with galaxy members of the same group. Our fossil record analysis gives evidences of a mutual interaction between objects \textbf{C}, \textbf{D}, \textbf{E} and \textbf{L}. These objects have a long SF burst that begins at the age of $\approx 6$ Gyr and start to decline at $\approx 3$ Gyr. They also have a weak SF burst that begins at $\approx 300$ Myr with a maximum at the age of $\approx 150$ Myr. This weak SF burst is not enough to reactivate their sSFR and is rapidly quenched. For the case of object \textbf{D}, the shared SF burst is only present at its outer region ($R>0.5R_{50}$), with an internal region, completely quenched after the age of $\approx 1$ Gyr. Their ASM show an outside-in assembly mode, with an average $T_{90}$ ($T_{50}$) age of 2.8 (5) Gyr (see cols 9 and 10 of Table \ref{tab2}). Their mass-weighted ages have an average value of 4.5 Gyr, while their light-weighted ages have an average of 3.5 Gyr. Their {\sc D4000} index have an average value of 1.9 dex. The mass and light-weighted stellar metallicities have an average value (0.08 dex), which is slightly supersolar (Table \ref{tab2} cols 6 and 7). 

\emph{Star formation objects}: This group contains objects \textbf{G}, \textbf{H} and objects \textbf{K} and \textbf{O}. We cannot detect a stellar spectra for objects \textbf{K} and \textbf{O}, and therefore we cannot estimate the sSFH and the ASM; however, we put them within this group due to their strong nebular emission lines for {[O\sc II]}$\lambda3728$, {H$\beta$}, {[O\sc III]}$\lambda\lambda4959,5007$. The sSFH  for objects \textbf{G} and \textbf{H} show the same initial SF burst at an age of $\approx 6$ Gyr. After this initial burst, their sSFR decreases until an age of $\approx 1$ Gyr, where the sSFR reaches a constant value ($\log sSFR\approx -10.0$ dex). The sSFR remains constant until an age of $\approx 200$ Myr, where objects \textbf{G} and \textbf{H} start to differentiate. Object \textbf{G} suffers a more steep decrease on their sSFR  reaching an observed time sSFR of -11.2 dex. On the other hand, object \textbf{H} maintains an slower decrement on their sSFR reaching a value of $\log sSFR\approx -10.7$ dex at the moment of the observation. The ASM of objects \textbf{G} and \textbf{H} show an inside-out assembly mode only for their internal regions ($R<0.5R_{50}$ vs $0.5R_{50}<R<R_{50}$).

Both objects have $T_{50}$ ages younger than 5.2 Gyr, with younger $T_{90}$ ages ($<1.5$ Gyr). Their light (mass) weighted ages have an average value of 0.9 (3.6) Gyr and their average light (mass) weighted stellar metallicities are -0.55 (-0.25) dex (see Table\ref{tab2}). The difference among the mass and light weighted metallicities are 0.2 and 0.4 dex for object \textbf{G} and \textbf{H}. The positive value of this difference indicates a possible metallicity dilution by the in-fall of pristine gas like objects \textbf{A} and \textbf{B} but in a much intense rate. The spectra of objects \textbf{G} and \textbf{H} (see Figure~\ref{plot_spectra2} and Table\ref{tab2}) show the emission of the nebular lines for {[O\sc II]}$\lambda3728$, {H$\delta$}, {H$\gamma$}, {H$\beta$}, and {[O\sc III]}$\lambda\lambda4959,5007$, with \textbf{H} displaying the most strong lines. These emission lines indicate the existence of an intense and resent SF burst. The stellar spectra show the existence of strong Balmer and {Ca} absorption lines. Their {\sc D4000} index have the lowest value of all our objects, with values of 1.47 and 1.3 dex for objects \textbf{G} and \textbf{H} respectively. For objects \textbf{K} and \textbf{O} we are only able to identify the emission of {[O\sc II]}$\lambda3728$, {H$\beta$}, {[O\sc III]}$\lambda\lambda4959,5007$. Due to the similarities of the nebular emission of objects \textbf{K} and \textbf{O} with objects \textbf{G} and \textbf{H}, we can speculate that objects \textbf{K} and \textbf{O} share a similar sSFR and SSP properties, indicating a recent and intense SF burst. 

\emph{Intermediate objects}: this group only contains object \textbf{J}, its spectrum is similar to the ones observed for the interacted objects. The SPS analysis reconstructs an sSFH that shows a late SF burst that reaches a maximum at the age of $\approx 6$ Gyr, with small SF bursts at $\approx2$ Gyr, $\approx500$ Myr, and $\approx80$ Myr. Those SF bursts are not strong enough to maintain the sSFR, continuously decreasing until the present day value of -13.5 dex. Their ASM shows an homogeneous mass growth at all radii showing a flat assembly mode at all epochs. The ASM gives a $T_{50}$ age of 5.3 Gyr with a $T_{90}$ age of 2.8 Gyr. The light (mass) weighted age is 3.6 (4.6) Gyr, with a {\sc D4000} index value of 1.82 dex. Its metallicity rounds the solar value (see Table\ref{tab2}) with a difference among the mass and light-weighted values of -0.03 dex.

\begin{figure}
	\includegraphics[width=0.5\columnwidth]{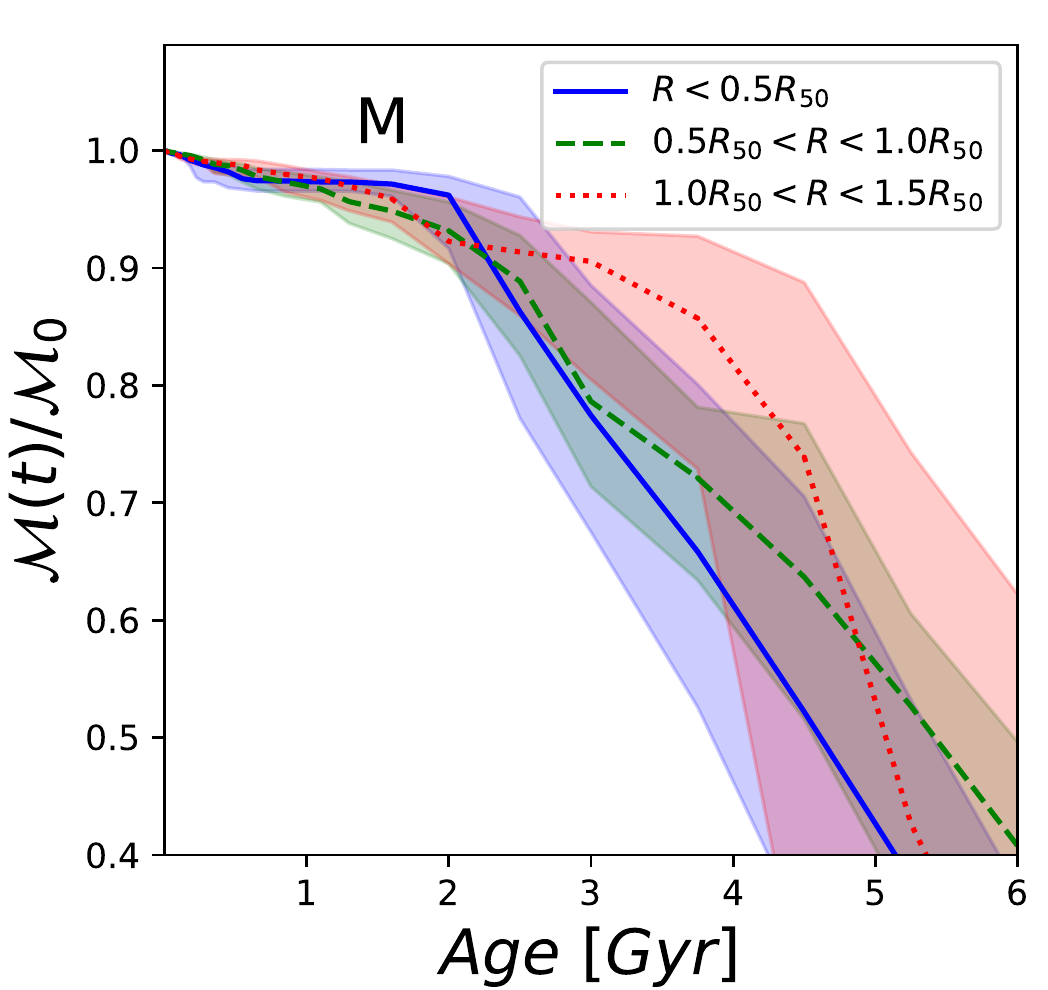} 
	\includegraphics[width=0.5\columnwidth]{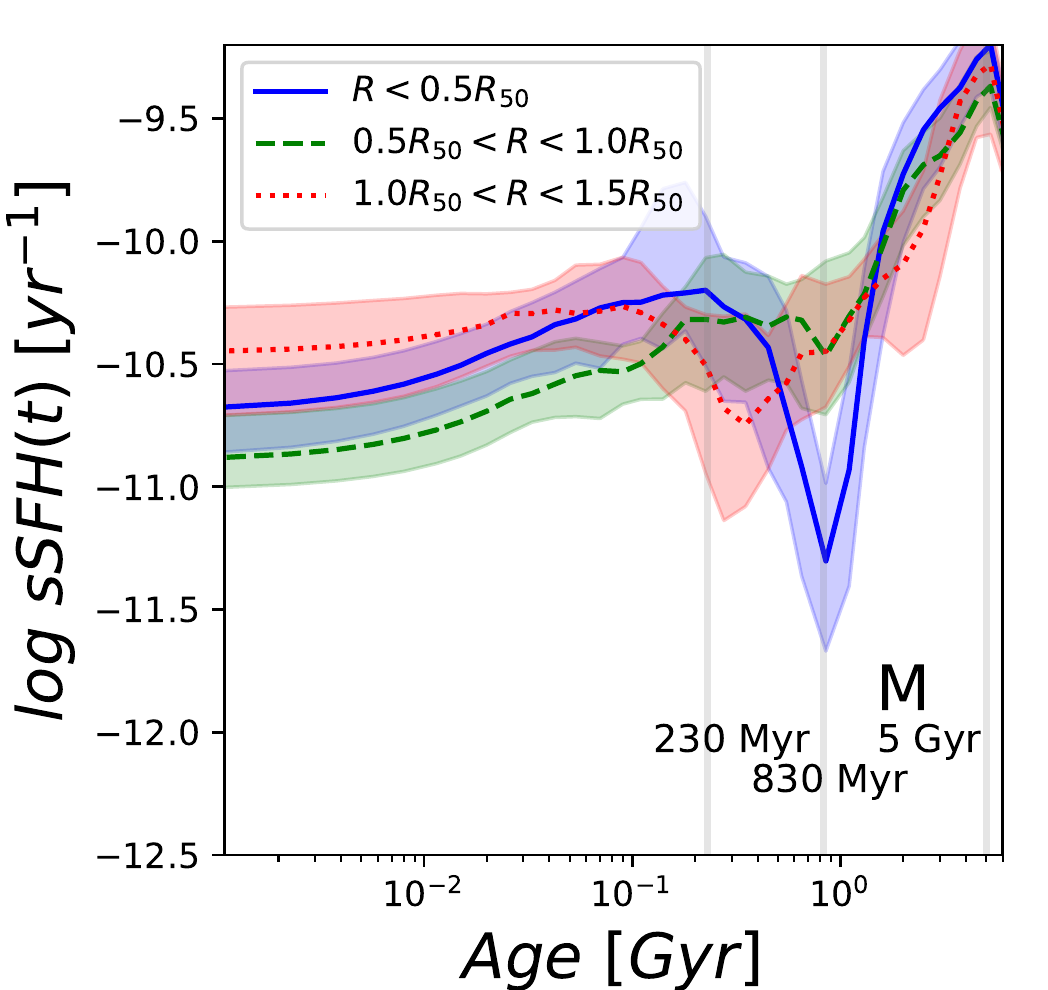} 
    \caption{Left panel: The ASM distributions for object $\mathbf{M}$ at three radial regions: $R < 0.5R_{50}$ (blue), $0.5R_{50} < R < R_{50}$ (green), and $R_{50} < R < 1.5R_{50}$ (red). Right panel: The sSFH for object \textbf{M} at the same radial bins and the same color code of the left panel. The shaded color areas are the $1\sigma$ variance for 100 iterations. The vertical gray lines represent the SF burst and quenching episodes described in the text.}
    \label{M_SFH}
\end{figure}

\subsection{Object M}

Figure~\ref{M_SFH} shows the fossil reconstruction of the sSFH and the ASM distribution of object \textbf{M}. The fossil reconstruction shows an initial star formation burst at the age of $\approx5$ Gyr, followed by a rapid SF quenching that drops the sSFR at a minimum value of $\approx-11.5$ dex at the age of $\approx830$ Myr. After this age, the sSFR is reactivated and reaches an sSFR peak ($\approx-10.2$ dex) at $\approx280$ Myr. Then, the sSFR slowly decreases until its present-day value (-10.8 dex). The spectrum of object $\mathbf{M}$ (see Figure~\ref{plot_spectra2}) is similar to the spectrum of object \textbf{G}, \textbf{H}, \textbf{K} and \textbf{O} showing a strong emission for the {[O\sc II]}$\lambda3728$ and {[Ne\sc III]}$\lambda3869$ lines. The very strong emission on the {[O\sc II]}$\lambda3728$ and {[Ne\sc III]}$\lambda3869$ nebular lines indicates the existence of an strong ionization source \citep{Osterbrock+2006} within object \textbf{M}. The ASM indicates an outside-in assembly mode, with a $T_{50}$ age of 5.2 Gyrs and a $T_{90}$ of 2.4 Gyr. The light (mass) weighted age of object \textbf{M} is 0.5 (4) Gyr while its metallicity has sub-solar values of -0.62 (-0.23) dex. Its {\sc D4000} index has a value of 1.3 dex and agrees with the measured light-weighed age.

\section{Analysis of J004150-091812}\label{group_ana}

\begin{figure}
	\includegraphics[width=\columnwidth]{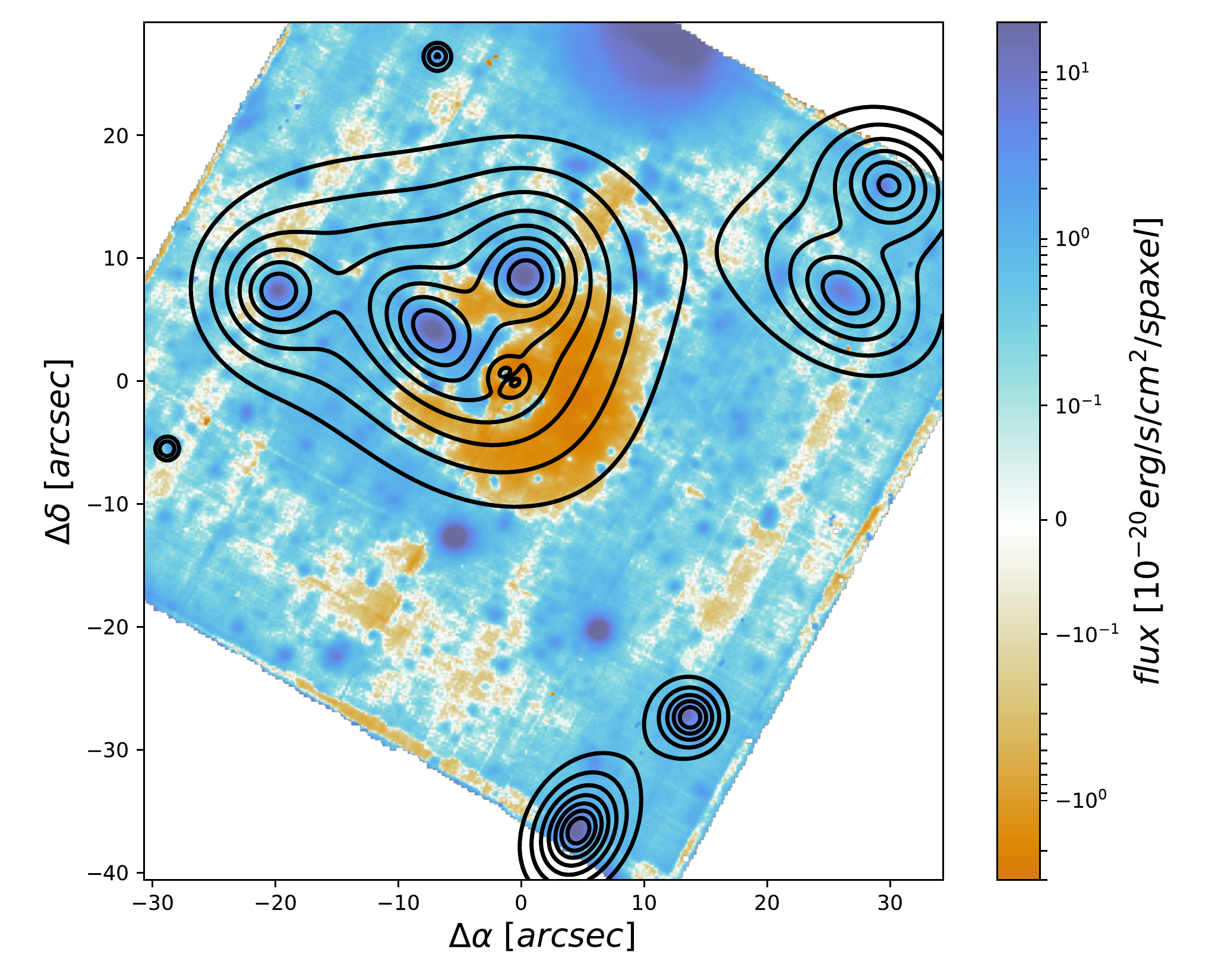}
    \caption{SB of the lensed images for each candidate member of J004150-091812. The black lines represent the SB isocontours of the images modeled by {\sc Lenstool}. The color map represents the residual map of Holm15A described in Figure~\ref{fig:Muse_map}. The color bar represents the observed SB flux.}
    \label{lenss_map}
\end{figure}

In this section, we explore the properties of the galaxy group J004150-091812 as a whole. We need to consider that J004150-091812 is behind of the massive halo of Holm15A, that contains one of the biggest SMBH \citep{Mehrgan+2019,Lopez-Cruz+2014}, and therefore the projected sky positions of objects \textbf{C}, \textbf{D}, \textbf{E}, \textbf{G}, \textbf{H}, \textbf{J}, \textbf{K}, \textbf{L} and \textbf{O} can be perturbed by the strong/weak lensing of the gravitational potential of Holm15A. To explore this possibility, we analyze the image distortion using the {\sc Lenstool} code \citep{Kneib+1996,Jullo+2007,Jullo+2009}. With this code, we can model and quantify the image distortion of the source and trace the amplification of the gravitational lens of Holm15A. 

\subsection{Lensing Modelling}

The observed SB distribution of Holm15A and the spectroscopic analysis of objects \textbf{A} and \textbf{B} show that a simple profile can model the total mass distribution of Holm15A. We show in the previous section that object \textbf{A} is not part of the cluster and have a total stellar mass of $5.8\times10^9$ $M_{\odot}$, and object \textbf{B} has a total stellar mass of $3.7\times10^9$ $M_{\odot}$. These masses are one order of magnitude lower than the most recent mass estimation ($4\times10^{10}$ $M_{\odot}$) of the SMBH in the center of Holm15A \citep{Mehrgan+2019}. Therefore, the mass contribution of objects \textbf{A} and \textbf{B} to the total potential model is negligible. Hence, we can model the total mass distribution using only one potential. For this aim, we use a dual pseudo isothermal elliptical mass distribution \citep[PIEMD][]{Limousin+2005,Eliasdottir+2007} that is defined as:

\begin{equation}
    \rho(r)=\frac{\rho_0}{(1+r^2/R_{core}^2)(1+r^2/R_{cut}^2)}
\end{equation}

We use the fitted values of our SB modeling of Holm15A to fix the ellipticity and position angle for the projected mass distribution. 

In addition, we use the mass profile measured by \citet{Mehrgan+2019} to constrain the values of $\rho_0$, $R_{cut}$ and $R_{core}$. \citet{Mehrgan+2019} estimate the mass profile by detailing modeling the stellar kinematics of Holm15A. We also model the mass contribution from the Holm15A SMBH to the total projected mass distribution as a Dirac delta function with a peak that equals the total mass of the SMBH measured by \citet{Mehrgan+2019}. For the lensing modeling, we consider all the galaxy members of J004150-091812 as single images each. We do not find any evidence of the existence of any strong lensed image that could form an arc image or multiple images from the same source. The absence of any strong or weak lensed image, will return in an unconstrained lens model. Therefore, it is a good choice to obtain the lens model by constraining the mass profile with the \citet{Mehrgan+2019} estimation. In Figure~\ref{lenss_map}, we show the best lensed SB model obtained with {\sc Lenstool}. It shows the models for the SB distributions for the 9 group galaxy candidate members and over-plotted the flux residuals of the SB fit of Holm15A.

\begin{figure}
	\includegraphics[width=\columnwidth]{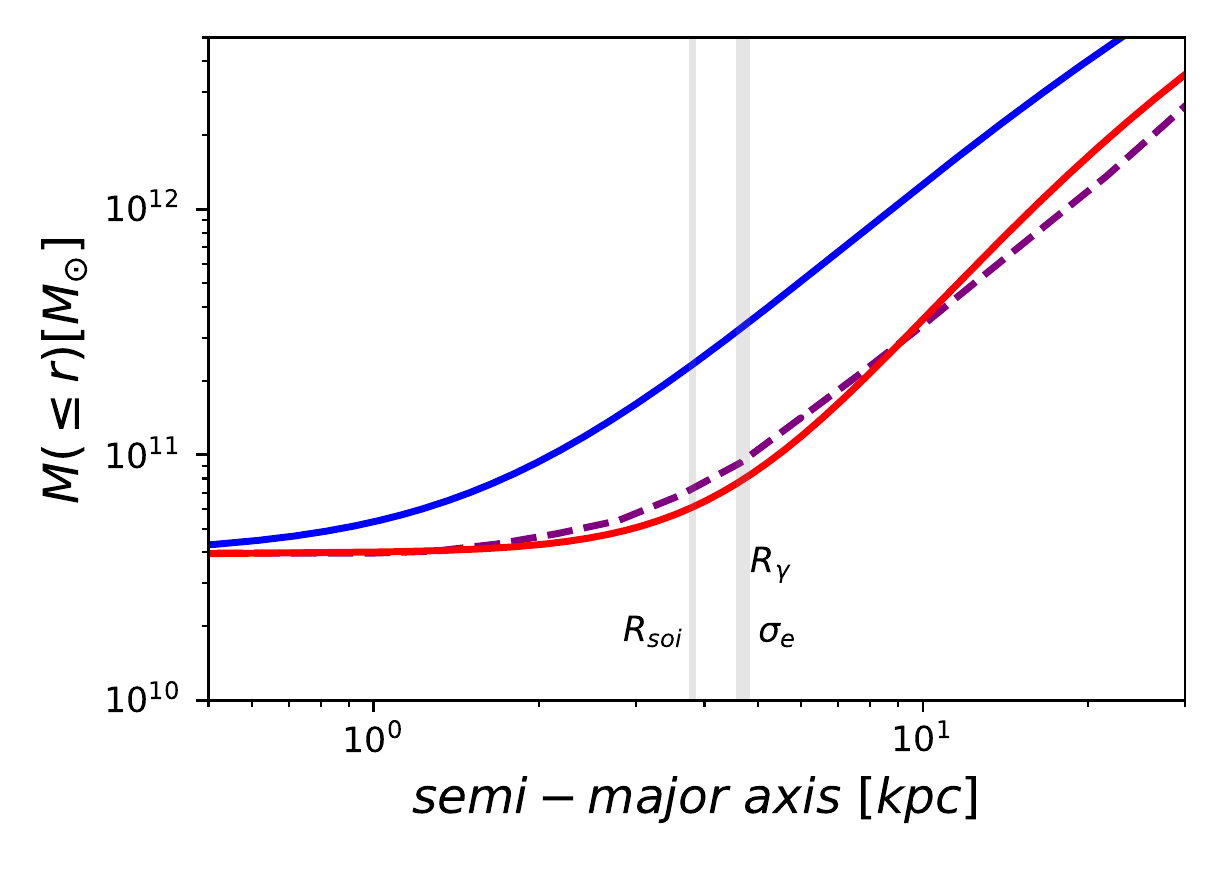}
    \caption{The cumulative mass profile of Holm15A is estimated with {\sc Lenstool}. The red line represents the three-dimensional enclosed mass profile. The blue line represents the two-dimensional projected cumulative mass profile. The purple dashed line profile represents the cumulative mass profile predicted by \citet{Mehrgan+2019}. The vertical gray lines represent the \citet{Mehrgan+2019} SOI scale, the $\sigma_e$ scale from the eccentricity profile of Holm15A, and the $R_{\gamma}$ scale from the Nkker profile fit of Holm15A.}
    \label{lenss_prof}
\end{figure}

The mass model returned by {\sc Lenstool} is a PIEMD potential with $R_{core}=$14.54 kpc, $R_{cut}=$1077.3 kpc and $\rho_0=9.5\times10^7$ $M_{\odot}$kpc$^3$. With these values, we can estimate the total enclosed mass at a given radius (see Figure~\ref{lenss_prof}). Following the analysis presented by \citet{Lopez-Cruz+2014}, we can use the cusp radii of our Nuker profile to estimate the size of the sphere of influence (SOI) of the SMBH. Therefore, the total mass contained within $R_{\gamma}$ equals to $7.7\times10^{10}$ $M_{\odot}$. On the other hand, the drop in the measured eccentricity of the isophotes in the inner radii of Holm15A (Figure~\ref{fig:fit_prof}) can be related to the dynamical mixing of the SMBH \citep{Lopez-Cruz+2014}. Therefore, the value of $\sigma_e$ could be used as a tracer of the SOI: its total mass contained within this scale is $8.1\times10^{10}$ $M_{\odot}$. Finally, using the value of SOI ($R_{soi}=3.8$ kpc) measured by \citet[][]{Mehrgan+2019}, we found a total enclosed mass of $6.1\times10^{10}\ M_{\odot}$. The PIEMD obtained from {\sc Lenstool} agrees with the total cumulative mass profile measured by \citet{Mehrgan+2019}. In this way, we can use the lens model derived by {\sc Lenstool} to correct the sky projected positions for the galaxy members of J004150-091812.
%


\begin{figure}
	\includegraphics[width=\columnwidth]{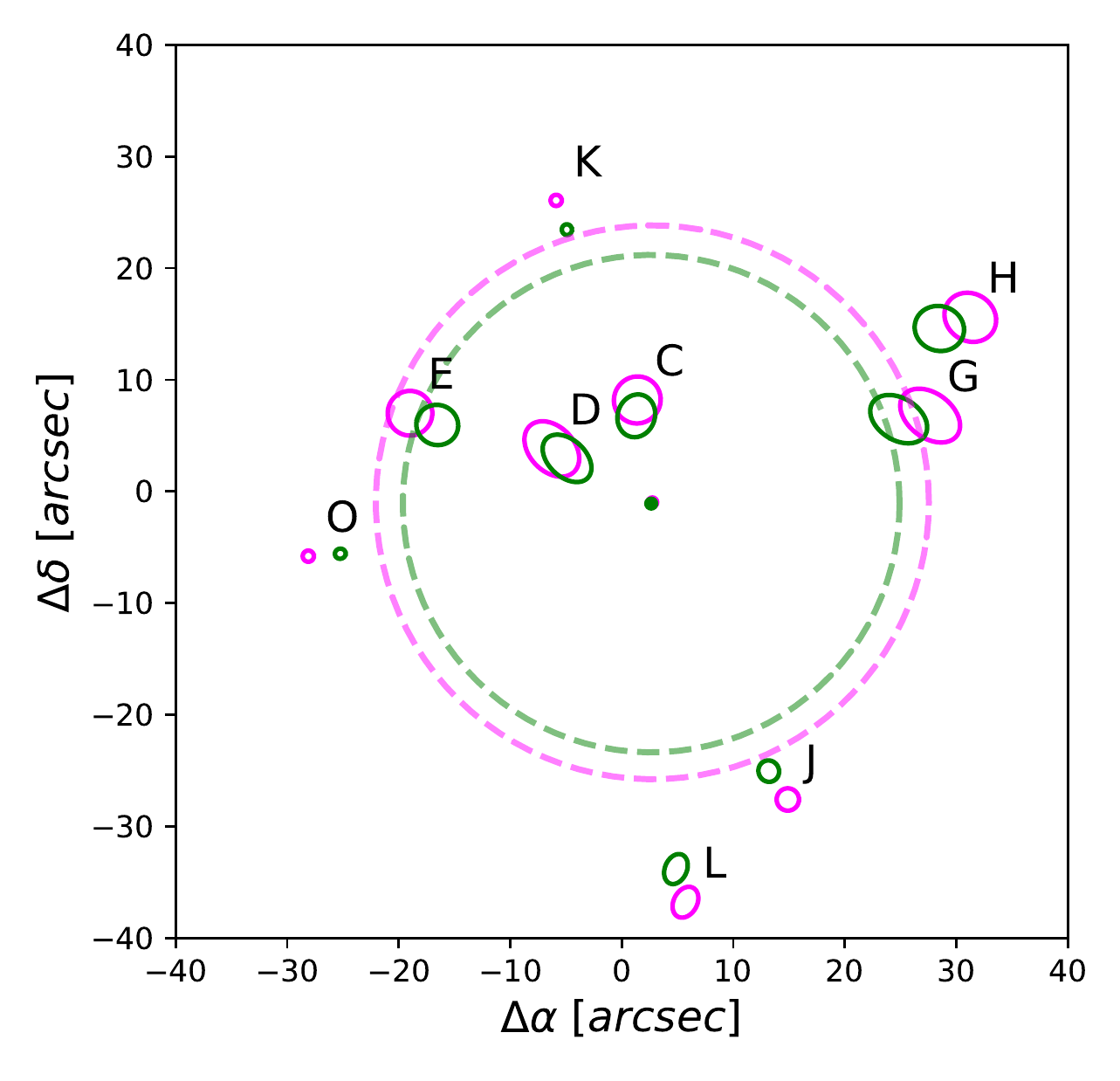}
    \caption{Sky projected positions of each galaxy member of J004150-091812. The size and shape of the objects represent the size of one $R_e$ with its measured eccentricity and inclination. The solid point represents the centroid of the galaxy group. The dashed circles represent the lower limit of the galaxy group scale $R_g$. The pink color represents the positions of the group members without the lensing correction of Holm15A. The green color represents the positions of the same objects with the lensing correction.}
    \label{group}
\end{figure}

\subsection{Scale of J004150-091812}

In Figure~\ref{group}, we show the positions of all group members of J004150-091812 with and without the perturbations of the gravitational lens of Holm15A. With the positions of the group members, we calculate the group centroid by weighting its positions by the logarithm of their stellar masses:

\begin{equation}
    \delta_{g}=\frac{\sum_i\log_{10}M_{\star,i}\delta_i}{\sum_i\log_{10}M_{\star,i}},
\end{equation}
\begin{equation}
    \alpha_{g}=\frac{\sum_i\log_{10}M_{\star,i}\alpha}{\sum_i\log_{10}M_{\star,i}}.
\end{equation}

For weighting the object $\mathbf{K}$ and $\mathbf{O}$, we use the lowest measured stellar mass of the group members as an upper limit of their stellar masses\footnote{This mass is the mass of the object $\mathbf{H}$ with $\leq 1.1\times10^{10}M_{\odot}$}. We calculate the group centroid with and without the correction of the gravitational lens. Both centroids are almost identical, with a small difference of $\approx0''.1$ in declination. The final centroid of the galaxy group is $\alpha=$00h41m50.4s, $\delta=$-09d18m11.86s.

With the centroid, we calculate the scale of the group ($R_g$). The FoV of the MUSE-WFM observation can bias this scale, and therefore we consider this scale as the lower limit of the real scale of the galaxy group. We calculate the scale of the group as the average distance from the centroid weighted by the stellar mass of each member:

\begin{equation}
    R_{g}\geq\frac{\sum_i\sqrt{(\delta_i-\delta_g)^2+(\alpha_i-\alpha_g)^2}\log_{10}M_{\star,i}}{\sum_i\log_{10}M_{\star,i}}.
\end{equation}

Considering the correction from the Holm15A lens, we obtain a value of $R_g=22''.3\pm0''.4$ or 146$\pm$3 kpc. Without this correction, we obtain a value of $R_g=24''.8\pm0''.4$ or 163$\pm$3 kpc (see Figure~\ref{group}).

\subsection{Dynamical Mass of J004150-091812}

From the measured redshift values of the galaxy members of J004150-091812, we measure its line-of-sight (LOS) velocity dispersion ($\sigma_{g}$). We estimate the 68.2 and 34.1 percentiles and determine the $1-\sigma$ LOS velocity value of the galaxy members. The final velocity dispersion value is $\sigma_{g}=622\pm300\ km/s$. With the scale and the velocity dispersion, we use the standard definition of the virial theorem \citep{Zwicky+37,Girardi+98} to estimate the dynamical mass of the group:

\begin{equation}
    M_{dyn}\geq A\frac{R_g\sigma_{g}^2}{G},
\end{equation}
where $A$ is a scaling factor that takes into account the geometry of the system and the dynamical state in order to estimate an unbiased value of the dynamical mass of the system \citep{Robotham+2011,Aquino+2020}. The value of A will be greater or equal to one, with $A=1$ as the ideal case. This study will express the dynamical mass in terms of A for simplicity. G is the gravitational constant in the corresponding units.\footnote{$4.3009\times10^{-6} kpc M_{\odot}^{-1}(km s^{-1})^2$}. Finally, we obtain that J004150-091812 has a lower limit of the dynamical mass of $M_{dyn}\geq 1.2\pm0.8\times10^{13} M_{\odot}A$, with the correction of the Holm15A lens. Without using the lens correction, we obtain a value of $M_{dyn}\geq 1.5\pm0.8\times10^{13} M_{\odot}A$. It is important to note that this mass estimation assumes that the J004150-091812 galaxy group is completely relaxed, however this assumption is not necessary true. We observe evidence that J004150-091812 is actually in an assembly process with at least three merger trees (see discussion below).

\section{Summary and Discussion}\label{discuss}

The principal aim of this study is the identification and characterization of the principal astronomical objects hidden behind the Holm15A SB. This characterization is a crucial step in separating structures that are physically linked within Holm15A from those that are external objects. This step will clean the path for a more depth study on the nature of the SMBH within Holm15A. For this aim, we use archive MUSE IFU observations of Holm15A and model and subtract its spectral contribution. Looking at the residuals, we detect 14 major structures from which we could extract their spectral information once cleaned from the Holm15A emission. We analyze the spectral information of all the objects and determine that ten objects present a well-defined galaxy spectra from which we can apply the fossil method to explore their past evolution and interlinkage among each other. From the spectral analysis, we find the following results:

\begin{itemize}
    \item The spectroscopic analysis confirms that all the 14 detected objects are not part of Holm15A. 
    
    \item Two objects (object \textbf{F} and \textbf{I}) are point sources that came from a quasar emission at z=1.5637 (object \textbf{F}) and a field star (object \textbf{I}).
    
    \item We detect two near objects on the vicinity of Holm15A (object \textbf{A} and \textbf{B}). Object \textbf{A} can be classified as a late-type galaxy with a well-defined bar structure. It has a redshift value of 0.0498, from which we can conclude that it is not dynamically linked to the  Abell 85 galaxy cluster. 
    Object \textbf{B} can be classified as an early-type galaxy with a redshift value of 0.0563, that confirms that it is a galaxy member of the Abell 85 cluster. 
    The SSP properties of both objects indicate the existence of two principal SSP components, one young (3 < Gyr) with subsolar metallicity and one old (4 > Gyr) with also subsolar metallicity but heavier than the young population. Therefore, the SSPs properties support the scenario of a possible rejuvenation of the SF activity with the dilutions of the gas content due to the in-fall of pristine gas \citep{Maiolino+2019,Artemi+2022}. This mechanism explains the halt in the SF quenching observed on both objects. 
    
    \item Object \textbf{M} is the farthest object that we detect with a well-defined galaxy spectrum. It has a redshift value of 1.0126; at that redshift, the universe had an age of 5.9 Gyrs. Therefore, the SPS analysis of object \textbf{M} allows us to access the archaeological information at the cosmic noon without suffering from a significant bias inherited from the archaeological inferences limitations \citep{Ibarra-Medel:2019aa}. 
    Their nebular emission lines confirm the archaeological view of a recent reactivation of the SFR. In addition, their SSP properties indicate a very strong dilution of the SSP metallicities between young and old populations with a difference of 0.39 dex. Its stellar spectra present a very well-defined stellar absorption line but it cannot be classified as a post-starburst (or A+E) galaxy due to the observed strong emission on {[O\sc II]}$\lambda3728$ \citep{Dressler+1983,Poggianti+1999,Yang+2008}.
    
    \item We detect nine objects that are possible candidates to be a galaxy group (J004150-091812) at a redshift of 0.5814. Of the nine objects, seven have well-defined galaxy spectra from which we obtain their archaeological reconstruction. We detect in four of them (objects \textbf{C}, \textbf{D}, \textbf{E} and \textbf{L}) the existence of shared star formation burst at the ages of 6-3 Gyrs and 300-150 Myrs. 
    These common SF burst episodes signal the possible scenario of past interactions among each other. It can be used to trace their crossing times within the galaxy group \citep{Lopez-Cruz+2019}. On the other hand, one object (object \textbf{J}) also presents a similar SFR shutdown as the past four interacted objects, but it has different SFR episodes at different epochs. However, the absence of a common SF burst did not imply that object \textbf{J} is not a galaxy member of J004150-091812, but rather it signals that it could be falling into the cluster in a later epoch. Therefore, the fossil inference suggests that object \textbf{J} can be a different merger tree on the assembling of J004150-091812 \citep{Tweed+2009}. In addition, we detect two galaxies with very different sSFHs (object \textbf{G} and \textbf{H}). These objects present a very active SF activity during its past and present.
    Their emission spectra confirms the previous by showing a very intense emission on the {[O\sc II]}$\lambda3728$, {H$\delta$}, {H$\gamma$}, {H$\beta$}, {[O\sc III]}$\lambda\lambda 4959,5007$ nebular lines. Their SPS properties also show evidence of the dilution of the SPS metallicities between the old and young stellar populations, indicating the existence of a mechanism that provides fresh gas that can feed and sustain the observed SFR. 
    The objects \textbf{K} and \textbf{O} did not have a well-defined stellar galaxy spectrum, and therefore we cannot apply our spectral analysis. On the other hand, we can detect an strong nebular emission that indicates a intense and recent star formation activity in a very similarly way as objects \textbf{G} and \textbf{H}. In the same direction as object \textbf{J}, the difference among the sSFH with the other members of the group didn't imply that objects \textbf{G}, \textbf{H}, \textbf{K} and \textbf{O} aren't members of J004150-091812. Objects \textbf{G}, and \textbf{H} (and possibly \textbf{K} and \textbf{O}) share a very similar sSFH with common sSFH episodes. Therefore, these galaxies could form a sub-group that recently falls into the gravitational potential of J004150-091812, showing another merger tree on the assembly history of J004150-091812. The stellar spectra of objects \textbf{G} and \textbf{H} show very intense and well-defined stellar absorption lines. However, these galaxies cannot be identified as E+A due to the intense emission of {[O\sc II]}$\lambda3728$. On the other hand, these galaxies can be the precursors of E+A galaxies, with a future SF quench after they completely join into the J004150-091812 galaxy group and suffers multiple galaxy mergers \citep{Bekki+2005}.
    
    \item We estimate the mass and size of J004150-091812 using the measured sky positions and redshifts for each galaxy member. To estimate the scale and the central position of J004150-091812, we first correct the sky positions of each member from the optical perturbations done by the gravitational lensing of the Holm15A potential. To perform such correction, we use the {\sc Lenstool} software. 
    With the lens correction, we estimate the scale of J004150-091812 as the stellar mass-weighted distance of each member to the centroid. Considering the lens correction, we find a scale of 146$\pm$3 kpc. Without this correction, we find a scale of 163$\pm$3 kpc. It is important to note that the FoV of the MUSE WFM biases these scales, and therefore, we only consider those scales as lower limits of the real scale of J004150-091812. 
    We find that J004150-091812 has a velocity dispersion of 622$\pm$300 km/s.
    Assuming that J004150-091812 is dynamical relaxed, its dynamical mass has a value of $M_{dyn}\approx 1.2\pm0.8\times10^{13} M_{\odot}A$ when we consider the lens correction and $M_{dyn}\approx 1.5\pm0.8\times10^{13} M_{\odot}A$ when we discard the lens correction. However, the SFHs of the galaxy members of J004150-091812 indicate that the galaxy group is in an assembly process and therefore it is not in a virial equilibrium. Following the mass-radius scale relationships explored by \citet{Onofrio+2020}, it is expected that a galaxy group with a measured mass of $\approx 10^{13} M_{\odot}$ would have a scale of $\approx 316$ kpc. For J004150-091812, we measure a scale of half than it is expected. The bias of the un-virialized stage of J004150-091812 overestimates its dynamical mass, explaining the differences in the galaxy group scales. 
    In addition, due to the limitations of the FoV on the scale estimation of J004150-091812, we are not available to classify J004150-091812 as a compact group according to the \citet{Hickson+1982} classical definition.  
    
    \item As a secondary result from this analysis, we propose using the circularization scale $\sigma_e$ of the ellipticity profile of Holm15A as an indirect estimation of the SMBH SOI scale. We will explore this possibility in a forthcoming paper.
\end{itemize}

\section{ACKNOWLEDGEMENTS}

We want to thank Castalia Alenka Negrete for her helpful comments and support in our manuscript preparation. We also thank Omar Lopez-Cruz for the valuable comments and suggestions on this work. This study has made use of the services of the ESO Science Archive Facility (program ID: 009.B-0193), the {\sc Lenstool} code \citep{Kneib+1996,Jullo+2007,Jullo+2009}, and {\sc PyQSOFit} code \citep{Guo+2018,Guo+2019,Shen+2019}. The paper has made use of the {\sc pyPipe3} code. We thank the IA-UNAM MaNGA team for creating this tool, and the CONACyT-180125 project for supporting them. 

\section{DATA AVAILABILITY}
The data underlying this article are available at the ESO public archive for the MUSE program ID 009.B-0193 with the following URL link: \url{https://archive.eso.org/wdb/wdb/eso/eso_archive_main/query?prog_id=099.B-0193}. We also put public available all the extracted datacubes for all our 14 objects at the next URL link: \url{https://github.com/hjibarram/Holm15A_objects_datacubes}. 

\bibliographystyle{mnras}
\bibliography{example} 

\appendix

\section{PSF modeling}\label{psf_ana}

\begin{figure}
	\includegraphics[width=\columnwidth]{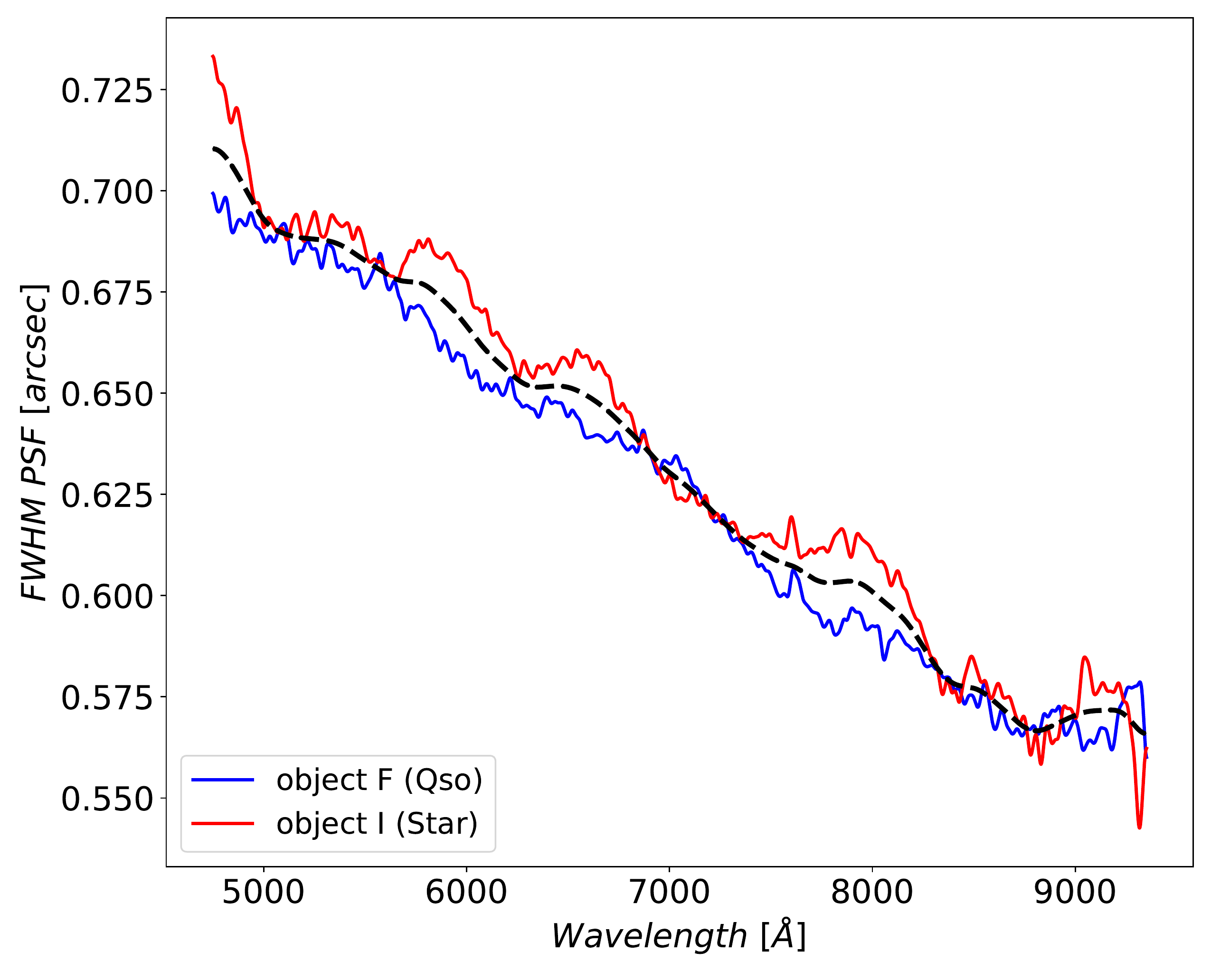}
    \caption{Measured FWHM PSF within the MUSE observation of Holm15A. The red dashed line represents the best fitted value of the PSF as a function of the wavelength using the SB profile of object \textbf{I}. The blue line represents the best fitted value of the PSF for object \textbf{F}. The black dashed line is the smoothed mean value of the two objects.}
    \label{psf_anaF}
\end{figure}

This appendix describes the measurement of the point spread function (PSF) and, therefore, the effective spatial resolution of the Holm15A MUSE WFM observation. For this aim, we use the extracted regions of objects \textbf{F} and \textbf{I} (see Figure\ref{fig:Muse_map}). We select each region to contain only the objects \textbf{F} and \textbf{I} without any other structure within it. The object \textbf{F} is a quasar at z=1.5637, and object \textbf{I} is a field star, and therefore are the ideal objects to measure the PSF within the MUSE datacube. We extract the flux maps for each spectral sampling of the datacube and fit a two-dimensional PSF SB profile. We assume that the PSF has an SB profile that follows a Moffat \citep{Moffat+1969,Trujillo+2001} shape with the following form:

\begin{equation}
    F(i,j)=A_t\times\left(1+\frac{(i-x_0)^2+(j-y_0)^2}{\alpha^2}\right)^{-\beta},
\end{equation}
where $A_t$ is the peak value of the PSF profile, $x_0$ and $y_0$ are their centroids, and $\alpha$ is the dispersion of the profile and $\beta$ traces the size of the "wings" of the PSF. As $\beta$ increases, the PSF "wings" decreases and the profile asymptotically tends to a Gaussian shape \citep{Trujillo+2001}.  During the fitting, we left as free parameters the values of At, $x_0$, $y_0$, $\alpha$ and $\beta$. We obtain the best fit values for the SB profile for objects \textbf{F} and \textbf{I} at each spectral value of the data cube. Finally, we estimate the PSF FWHM as $2\alpha\times\sqrt{2^{1/\beta}-1}$. In Figure~\ref{psf_anaF}, we plot the final measured values of the PSF within the MUSE data cube in terms of the wavelength.

We find that the value of $\beta$ remains constant thought the wavelength axis with a value of $\beta=2.5\pm0.2$ dex. The MUSE PSF contains a more extended wing size in comparison of a Gaussian PSF.

\section{Spectral Model of object F}\label{qsr_t}

\begin{figure}
	\includegraphics[width=\columnwidth]{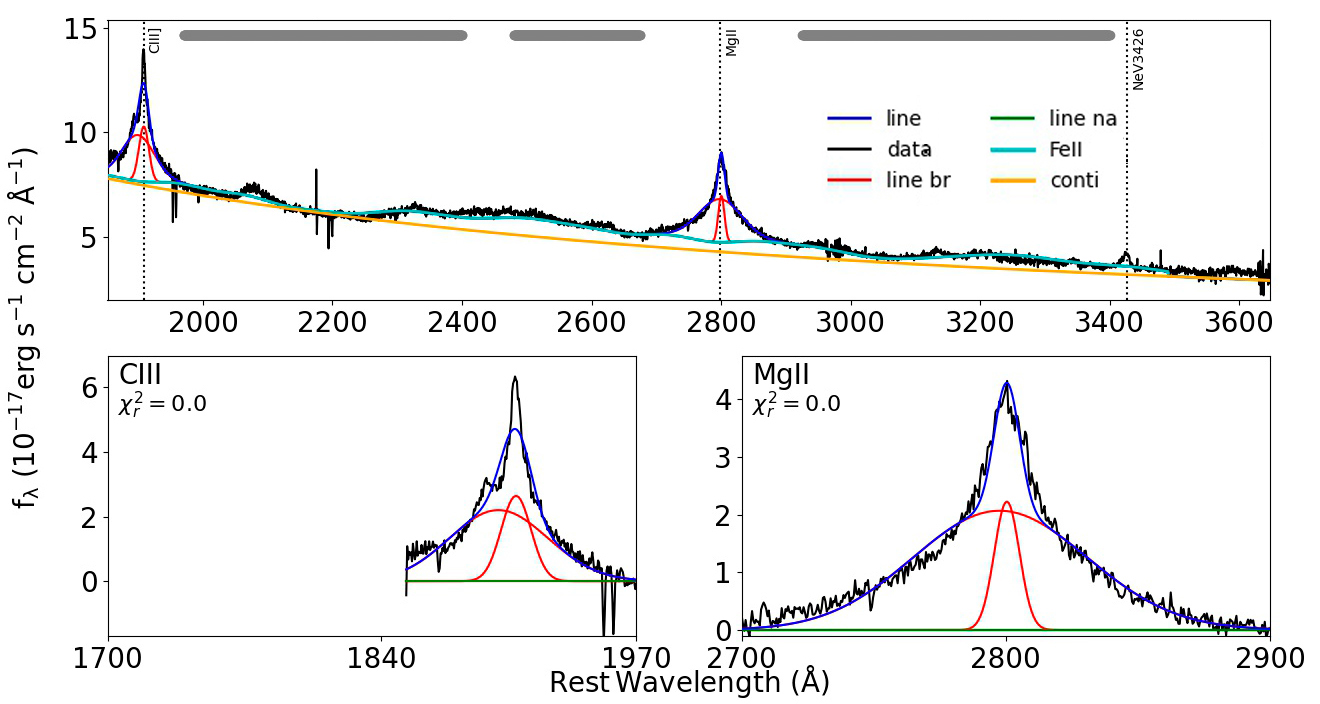}
    \caption{PSF spectrum model of object \textbf{F}. The plot was obtained from the {\sc PyQSOFit} code. The upper panel shows the input PSF spectrum of object \textbf{F}: the solid yellow line shows the continuum power law of the quasar, the blue line shows the fitted {Fe\sc II} emission, and the solid red line shows the fit of the broad component of {[C\sc III]}$\lambda1906$ and {Mg\sc II}$\lambda2800$ emission lines, the solid green line represents the narrow line component for the same emission lines, and the solid blue line represents the total contribution of all the fitted models. The left lower panel shows a zoom-in of the {[C\sc III]}$\lambda1906$ spectral region. The right lower panel shows a zoom-in of the {Mg\sc II}$\lambda2800$ spectral region.}
    \label{qsr_ana}
\end{figure}

With the PSF model, we can reconstruct the SB model cube for object \textbf{F}, to extract the PSF flux spectrum. The PSF flux spectrum is the integral of the PSF SB model at each spectral point and is defined as: $F_t=A_t\frac{\pi\alpha^2}{\beta-1}$. With the extracted PSF spectrum, we proceed to model the quasar spectrum of object \textbf{F}. To model the quasar spectrum , we use the {\sc PyQSOFit} tool\footnote{\url{https://github.com/legolason/PyQSOFit}} \citep{Guo+2018,Guo+2019,Shen+2019}. This tool fits the galaxy host continuum, the power-law contribution, the {Fe\sc II} fluxes, and fits the broad and narrow components from the quasar emission lines \citep{Vanden+2001}. 

In Figure \ref{qsr_ana}, we show the PSF quasar spectrum for object \textbf{F}, with the fitted models delivered by {\sc PyQSOFit}. We measure a quasar power-law with a slope of -1.54 dex, with an FWHM for the {Mg\sc II}$\lambda2800$ emission line broad component of 2,800 km/s. We also measure the continuum flux at $\lambda L_{3000}=21\times 10^{44}$ $erg/s$. Therefore, using the {Mg\sc II} estimator for the black hole (BH) mass \citep{Wang+2009}, we measure a BH mass of $1.3\times10^9 M_{\odot}$. With the PSF quasar spectrum model, we obtain the map of residuals from the extracted region of object \textbf{F}. From the residual map, we cannot find any structure associated with the host galaxy of object \textbf{F}. Hence the spectrum observed for object \textbf{F} is completely dominated by the quasar emission. 

\bsp	
\label{lastpage}
\end{document}